\documentclass[twocolumn]{aastex62}

\usepackage{graphics} 
\usepackage{hyperref}
\usepackage{epsfig} 
\usepackage{mathptmx} 
\usepackage{times} 
\usepackage{amsmath} 
\usepackage{amssymb}  
\usepackage{wasysym}
\usepackage{comment} 
\usepackage{lipsum} 
\usepackage{graphicx}
\usepackage{longtable}
\usepackage[normalem]{ulem}

\newcommand\msun{M$_\odot$}

\shorttitle{
Black Hole Candidates in the Microlensing Events M96-B5 and M98-B6
}
\shortauthors{Abdurrahman et al.}

\begin{document}
\title{On the Possibility of Stellar Lenses in the Black Hole Candidate Microlensing Events MACHO-96-BLG-5 and MACHO-98-BLG-6}

\author[0000-0002-9915-8195]{Fatima N. Abdurrahman}
\affiliation{Department of Astronomy, University of California, Berkeley, CA, USA 94720}

\author{Haynes F. Stephens}
\affiliation{Department of Geophysical Sciences, University of Chicago, Chicago, IL, USA 60637}

\author[0000-0001-9611-0009]{Jessica R. Lu}
\affiliation{Department of Astronomy, University of California, Berkeley, CA, USA 94720}

\begin{abstract}
 Though stellar-mass black holes (BHs) are likely abundant in the Milky Way (N$=10^8$--10$^9$), only ~20 have been detected to date, all in accreting binary systems \citep{casares2006observational}. Gravitational microlensing is a proposed technique to search for \textit{isolated} BHs, which to date have not been detected. Two microlensing events, MACHO-1996-BLG-5 (M96-B5) and MACHO-1998-BLG-6 (M98-B6), initially observed near the lens-source minimum angular separation in 1996 and 1998, respectively, have long Einstein crossing times ($>300$ days), identifying the lenses as candidate black holes. Twenty years have elapsed since the time of lens-source closest approach for each of these events, indicating that if the lens and source are both luminous, and if their relative proper motion is sufficiently large, the two components should be spatially resolvable. We attempt to eliminate the possibility of a stellar lens for these events by: (1) using Keck near-infrared adaptive optics images to search for a potentially now-resolved, luminous lens; and (2) examining multi-band photometry of the source to search for flux contributions from a potentially unresolved, luminous lens.

We combine detection limits from NIRC2 images with light curve data to eliminate all non-BH lenses for relative lens-source proper motions above 0.81 mas/yr for M96-B5 and 2.48 mas/yr for M98-B6. Further, we use WFPC2 broadband images to eliminate the possibility of stellar lenses at any proper motion. We present the narrow range of non-BH possibilities allowed by our varied analyses. Finally, we suggest future observations that would constrain the remaining parameter space with the methods developed in this work.

\end{abstract}



\section{Introduction} \label{sec:intro}
Detecting isolated black holes (BHs) remains a problem of both great importance and difficulty in astrophysics. Core-collapse supernova events mark the deaths of high-mass ($\gtrsim 8$ \msun) stars and are predicted to leave remnant BHs on the order of several to tens of $M_{\astrosun}$. An estimated  $10^{8}-10^{9}$ stellar-mass black holes are predicted to occupy the Milky Way \citep{agolandkami}. However, only $\sim$20 have been detected,  all in accreting binaries \citep{2007IAUS..238....3C,2014SSRv..183..223C,2013ApJ...769...16R}.  
More recently, BH binaries have been identified by gravitational waves emitted from their mergers \citep{PhysRevLett.116.061102}. 
Isolated BHs, which could comprise the majority of the BH population \citep{wiktorowicz2019populations}, remain elusive, with no confirmed detections to date. Detecting isolated BHs and measuring their masses would help constrain the number density of BHs as well as the initial-final mass relation--which designates which stars become BHs versus neutron stars or white dwarves--in turn informing the understanding of BH formation, supernova physics, and the equation of state for nuclear matter \citep{2016ApJ...830...41L}. 

While an isolated BH does not produce a detectable electromagnetic radiation signature, it is in principle detectable as a lens in a gravitational microlensing event, whereby the gravitational potential of a massive lensing object refracts and focuses the light of a background source in accordance with their time-dependent angular separation on the plane of the sky \citep{1986ApJ...301..503P,1993PhyW....6...26P,1674-4527-12-8-005}. One characterizing feature of such an event is the Einstein radius \textbf{$\theta_{\textrm{E}}$}, which describes the radius of the source image given the physical alignment of an event's source, lens, and observer, and is defined as:
\begin{equation}
\theta_{\textrm{E}}=\sqrt{\frac{4GM}{c^2}(d_L^{-1}-d_S^{-1})}
\end{equation}
where G is the gravitational constant, M is the lens mass, c is the speed of light, $d_L$ is the observer-lens distance and $d_S$ is the observer-source distance. More readily recoverable from a microlensing light curve, however, is the Einstein crossing time $t_{\textrm{E}}$, which is related to an event's Einstein radius and relative source-lens proper motion \textbf{$\mu_{rel}$} by:
\begin{equation}
\vec{\theta_{\textrm{E}}}=\vec{\mu_{\textrm{rel}}}t_{\textrm{E}}
\label{eq:tE}
\end{equation}
While $t_{\textrm{E}}$ encodes all of the physical parameters of an event (namely, lens mass and lens and source distances,) in the absence of additional signals such as finite source effects, astrometric shifts, or parallax, one is limited to applying a bayesian prior from a Galactic model and estimating (rather than measuring) the lens mass. The only second order effect considered in this work is parallax, which causes an asymmetric distortion to a point-source point-lens (PSPL) microlensing light curve. In microlensing formalism, the `microlensing parallax’, $\pi_{\textrm{E}}$, is defined as:

\begin{equation}
\vec{\pi_{\textrm{E}}} = \frac{1\: \textrm{AU}(d_L^{-1}-d_S^{-1})}{\theta_{\textrm{E}}}\hat{\theta_{\textrm{E}}}
\label{eq:piE}
\end{equation}

The microlensing parallax can be understood as the ratio between the Earth's orbit and the Einstein radius of the microlensing event, projected onto the observer plane.

Currently, several surveys monitor many tens of square degrees near the Galactic Bulge in search of the photometric variability characteristic of microlensing events — the fourth phase of the Optical Gravitational Lensing Experiment (OGLE-IV; \cite{udalski1993optical}, \cite{udalski2015ogle}), the Microlensing Observations in Astrophysics collaboration (MOA-II; \cite{bond2001real}, \cite{sako2008moa}), and the Korea Microlensing Telescope Network (KMTNet; \cite{kim2016kmtnet}). Historically, the MACHO Project\footnote{A full description of the MACHO Project can be found at \url{http://wwwmacho.anu.edu.au/Project/Overview/status.html}} used microlening to search specifically for MAssive Compact Halo Objects (MACHOs), a hypothesized form of dark matter in the Milky Way halo that could include BHs \citep{1993Natur.365..621A}. Previous works have considered events from the MACHO survey as well as the OGLE survey for the possibility of BH lenses \citep{wyrzykowski2016black, mao2002optical}, though they have largely been limited to the examination of microlensing light curves. In this work, we will similarly examine the light curves of two events discovered in the MACHO survey, with the addition of more recent data from several different telescopes. 

MACHO-96-BLG-5 and MACHO-98-BLG-6 (hereafter referred to as M96-B5 and M98-B6, respectively) are microlensing events first detected and observed using long-term photometric monitoring \citep{2002Ben}. M96-B5 and M98-B6 were two of forty-five candidates detected toward the Galactic bulge. M96-B5 was previously identified in 1996 as a BH candidate, due to its exceptionally long Einstein crossing time of $\sim$970 days. For this event, the mass of the lens was estimated to be $M = 6_{-3}^{+10} $ \msun, and its Heliocentric distance within the range of 0.5--2 kpc \citep{2006ApJ...651.1092N}. However, because these estimations were derived from fitting light curves alone--a process that necessarily exhibits degeneracies--they are heavily influenced by the imposed Galactic model prior. Another analysis of this event strongly excluded the possibility of the lens being a main-sequence star because of brightness constraints given by Hubble Space Telescope data, and their mass constraints ruled out the possibility of the lens being a neutron star \citep{2002Ben}. The event was found by \cite{2005MNRAS.361..128S} to be inconsistent with a microlensing event occurring in a galaxy that did not include stellar remnant populations, based on simulations of microlensing light curves created from Galactic models that did not include neutron stars, white dwarfs, or black holes.
M96-B5 was found to be measurably affected by microlensing parallax \citep{2002Ben}, and additionally exhibits minor perturbations due to xallarap \citep{2005ApJ...633..914P}, which refers to the accelerated motion of the source due to its binary companion, and which can complicate the determination of the source angular position at the time of the event \citep{0004-637X-784-1-64}. M96-B5 was designated to be a marginal BH candidate, with a determined $37\%$ likelihood of being a black hole \citep{2005ApJ...633..914P}. 

The second candidate in this work, M98-B6, is a microlensing event identified in 1998 as a possible BH. The mass of the BH candidate for the event is estimated to be $M = 6_{-3}^{+7}$ \msun \citep{2006ApJ...651.1092N}, though again, this result was dependent on Galactic models. The source star has a heliocentric radial velocity of $-39\pm20$ km/s and is classified as a G5 IV spectral type \citep{Soto:2007}. \cite{2002Ben} were not able to strongly exclude the possibility of the lens being either a main-sequence star or a neutron star from their constraints, as both possibilities still have small likelihoods due to the confidence levels used. The event is consistent with the simulations of \cite{2005MNRAS.361..128S}, as described in the previous paragraph, which would make the possibility of a main-sequence lens seem more likely. M98-B6 is also affected by parallax, with minor perturbations from xallarap \citep{2005ApJ...633..914P}. It is considered a weak BH candidate ($2.2\%$ likelihood) \citep{2005ApJ...633..914P}. 

The paper is structured as follows: In Section \ref{sec:data} we describe the observations and data sets that went into this study. In Section \ref{sec:reduction} we lay out the reduction processes used on the relevant data sets. In Section \ref{sec:limits} we examine the detection limits for a luminous lens in recent, near-infrared (NIR) images. In Section \ref{sec:lightcurves} we fit microlensing light curve models to each event and use the resultant fits to separately constrain the flux contributions from the lens and source. In Section \ref{sec:sources} we photometrically and astrometrically analyze the source stars in an effort to determine whether the lens is detectably luminous. In Section \ref{sec:results} we synthesize the different constraints, and then discuss the likelihood of each event being a BH in Section \ref{sec:discussion}. Finally, in Section \ref{sec:conclusion} we present the conclusions of our study.
\newline
\section{Data Sets} \label{sec:data}

\subsection{Keck Observations} \label{sec:keck}

\begin{deluxetable*}{ccccccccccc}

	\tabletypesize{\footnotesize}
	\tablecolumns{11} 
	\tablewidth{0pt}
	\tablecaption{NIRC2 Observations}
	\tablehead{
		\colhead{Target} \vspace{-0.1cm} & RA (J2000) & Dec (J2000)& Epoch (UT) & N$_{exp}$  & Int. Time (sec)  & Coadds & N$_{stars}$ & Strehl & FWHM (mas) & m$_{base}$  
	}
	\startdata
	M96-B5 & 18:05:02.5 & -27:42:17 & 2016-07-14 & 9   & 60 & 1 & 297 & 0.22 &  70 & 16.864$\pm$0.035 \\
	 & &  & 2017-05-21 &  5  & 5  & 6 & 130 & 0.21 & 67 & 16.909$\pm$0.032 \\
	M98-B6 & 17:57:32.8 & -28:42:45 & 2016-07-14  & 22 & 10 & 1 & 344 & 0.33 & 58 & 12.895$\pm$0.033 \\
	 & & & 2017-06-08  & 15 & 2   & 1 & 74 & 0.19 & 82 & 12.868$\pm$0.035
	\enddata
	\tablenotetext{}{Strehl and FWHM are the average values of all N$_{stars}$ stars over all N$_{exp}$ individual exposures. Baseline (unlensed) magnitude $m_{base}$ indicates the target magnitude in NIRC2's `Kp' filter.}
	\label{tab:nirc2}
\end{deluxetable*}

\begin{figure*}[t!]
	\centering
	\includegraphics[width=\textwidth]{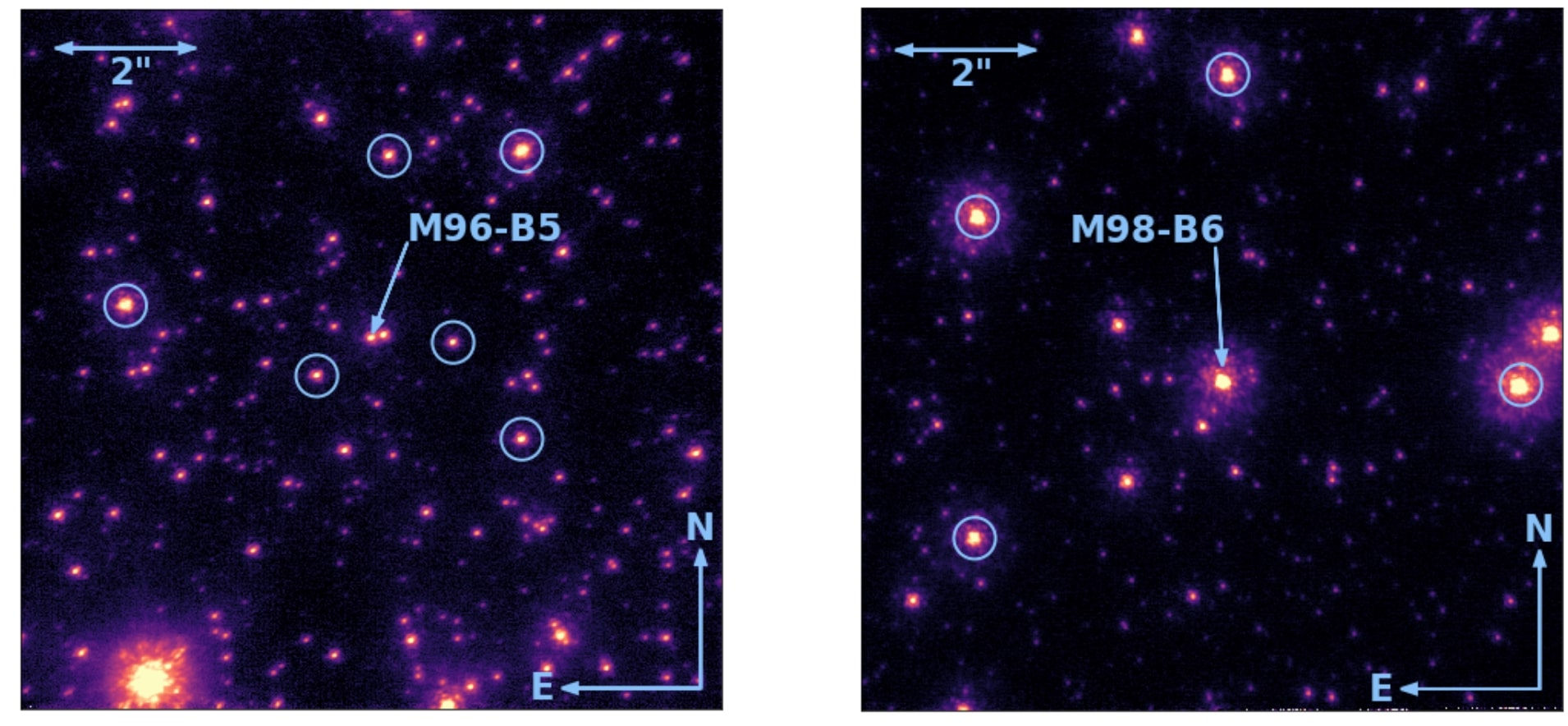}
	\caption{ 2016 NIRC2 images of M96-B5 {\em (left)} and M98-B6 {\em (right)}.	In each case, the background source star of the microlensing event is labeled with the event name. Stars used to create the mean PSF (cf. Section \ref{sec:plants}) are circled in light blue. PSF star selection was based on cuts in brightness, proximity to the target, and isolation. Note, the images have a logarithmic color scale.}
    \label{fig:frames}
\end{figure*}

The primary data presented in this paper are imaging observations of M96-B5 and M98-B6 taken with the NIR camera (NIRC2) on the W.M. Keck II 10m telescope behind  
the laser guide star adaptive optics (LGS AO) system \citep{2006PASP..118..297W}. 
Images were taken with the NIRC2 narrow camera in the Kp filter over a 10''$\times$10'' field with a plate scale of 9.952 mas pixel$^{-1}$ \citep{2016PASP..128i5004S}. Figure \ref{fig:frames} shows the two fields, in each case with the target marked at its center.

Both targets were observed on 2016 July 14 UT, approximately 20 years after peak magnification of the corresponding microlensing events. Each target was observed several times in succession with a total integration time of 540 s for M96-B5 and 220 s for M98-B6. In order to remove detector artifacts such as bad pixels, images were taken with a random dither pattern within a 0\farcs7 x 0\farcs7 box.

A second epoch of NIRC2 data was taken for each of the two events in 2017 to constrain motions. Additional details of all observations are provided in Table \ref{tab:nirc2}.

\subsection{MACHO Light Curves} \label{sec:macho}
 Both events discussed in this paper were first identified by the MACHO Project, in which 10-20 million stars in the Galactic bulge were surveyed from 1993-1999. The primary data set comes from the Mt. Stromlo 1.3m telescope in two filters: $B_{\textrm{MACHO}}$ ($\sim$450--630 nm) and $R_{\textrm{MACHO}}$ ($\sim$630--760 nm). The establishment of the MACHO Alert system made follow up observations from additional telescopes possible. As such, there are additional data for both events from the CTIO 0.9m telescope and the Mt. Stromlo 1.9m telescope \citep{alcock2000macho}.
 
In this work, we used the reduced and calibrated light curves for these targets from \citep{2002Ben}.

\subsection{Gaia Early Data Release 3} \label{sec:gaia}
In order to define an absolute astrometric reference frame for measurements of the targets' proper motions, we use data from Gaia Early Data Release 3 (EDR3)  \citep{brown2020gaia, lindegren2020gaia, riello2020gaia, fabricius2020gaia}. The EDR3 source list was queried for all sources within a 36''$\times$36'' box centered on each target. For the stars yielded by this search, we used only position information (RA and Dec). 
Additionally, a star in EDR3 was found to correspond to the location of the event M98-B5, the relevant characteristics of which are given below in Table \ref{tab:gaia}. This star is likely the source in the lensing event, though it could be the lens and source together (a possibility we will further examine in Sections \ref{sec:flens} and \ref{sec:sources}). 

\begin{deluxetable}{lll}[h]
    \tabletypesize{\footnotesize}
    \tablewidth{0pt}
    \tablecaption{Gaia Data for M98-B5 Position Match}
    \tablehead{
    Quantity  & Value & Error 
    }
    \startdata
    Source ID & 4062585513471740288 &  \\
    RA (deg) & 269.38573681 & 2.4$\times10^{-8}$ \\
    Dec (deg) & -28.71091478 & 1.9$\times10^{-8}$ \\
    RA Proper Motion (mas/yr) & -2.98 & 0.11 \\
    Dec Proper Motion (mas/yr) & 0.812 & 0.075 \\
    Parallax (mas) & 0.150 & 0.096 \\
    Magnitude (g) & 16.9661 & 0.0024 \\
    T$_{\textrm{eff}}$ (K) & 4465.25 & +478.69/-582.60 \\
    ruwe & 1.29 & \\
    \enddata
    \tablenotetext{}{Properties of the Gaia EDR3 object corresponding to the location of the event M98-B5.}
    \label{tab:gaia}
\end{deluxetable}

\bigskip
\subsection{HST Broadband Photometry} \label{sec:hst}
In order to compile multi-band photometry for the microlensing event sources, we utilized archival data from the Hubble Space Telescope’s WFPC2 instrument (HST-GO-8654, PI: Bennet, David P.). Observations were taken across 7 epochs between 1999 and 2003 for M96-B5, and a single epoch in 2000 for M98-B6. The data were taken in four wide filters: F439W, F555W, F675W, and F814W. In all cases, the target fell on the PC chip of the 4-chip camera. This 800$\times$800 pixel CCD has a plate scale of 45.5 mas pixel$^{-1}$, corresponding to a 36''$\times$36'' field of view \citep{gonzaga_biretta__2010}. Additional details on the HST images used are shown in Table \ref{tab:hst}. The data used were downloaded in 2018 November. 

\begin{deluxetable}{lccccc}
\tabletypesize{\footnotesize}
\tablewidth{0pt}
\tablecaption{WFPC2 Observations}
\tablehead{
\colhead{Target} \vspace{-0.1cm}  & Filter & Epoch & $N_{exp}$ & $t_{int}$ (s) & $m_{base}$ 
}
\startdata
M96-B5 & F439W & 2000-06-11 & 6 & 2200 & 20.609$\pm$0.034  \\
\hline
 & F555W & 1999-06-15 & 2 & 800 & 18.826$\pm$0.016  \\
 &  & 2000-06-11 & 5 & 3240 & 19.027$\pm$0.027  \\
 &  & 2001-06-03 & 2 & 800 & 18.877$\pm$0.095  \\
 &  & 2001-10-01 & 2 & 800 & 18.884$\pm$0.047  \\
 &  & 2002-05-25 & 2 & 800 & 19.021$\pm$0.009 \\
 &  & 2002-10-02 & 2 & 800 & 18.995$\pm$0.041  \\
 &  & 2003-05-27 & 2 & 800 & 19.042$\pm$0.009  \\
 \hline
 & F675W & 2000-06-11 & 5 & 1080 & 17.969$\pm$0.015 \\
 \hline
 & F814W & 1999-06-15 & 4 & 800 & 17.141$\pm$0.054  \\
 &  & 2000-06-11 & 5 & 3240 & 17.304$\pm$0.017 \\
 &  & 2001-06-03 & 4 & 800 & 17.336$\pm$0.015  \\
 &  & 2001-10-02 & 4 & 800 & 17.310$\pm$0.026  \\
 &  & 2002-05-25 & 4 & 800 & 17.333$\pm$0.020  \\
 &  & 2002-10-02 & 4 & 800 & 17.365$\pm$0.008 \\
 &  & 2003-05-27 & 4 & 800 & 17.341$\pm$0.009  \\
 \hline \hline
M98-B6 & F439W & 2000-06-23 & 1 & 40 & 17.996$\pm$0.044  \\
\hline
 & F555W & 2000-06-23 & 1 & 260 & 16.038$\pm$0.044  \\
 \hline
 & F675W & 2000-06-23 & 1 & 100 & 14.677$\pm$0.044  \\
 \hline
 & F814W & 2000-06-23 & 1 & 100 & 13.834$\pm$0.044  \\
\enddata 
\tablenotetext{}{ Observational parameters for archival HST data of MB96 and MB98. For a given target, filter, and epoch, there are $N_{exp}$ frames with a combined integration time of $t_{int}$. The baseline (unlesnsed) magnitude $m_{base}$ is the mean value and uncertainty in the mean of all frames in epochs with multiple observations. For epochs with a single observation, the uncertainty was set by inflating the uncertainties of multi-frame epochs with similar exposure times by the squareroot of the number of frames. }
\label{tab:hst}
\end{deluxetable}

\section{Raw Reduction and Catalog Creation} \label{sec:reduction}

\begin{figure*}[t]
	\centering
	\includegraphics[width=\textwidth]{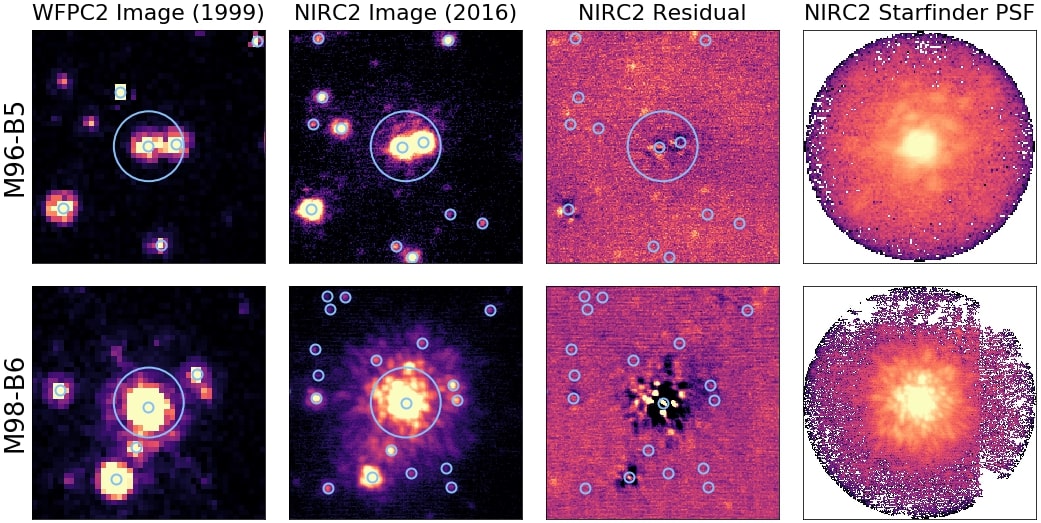}
	\caption{ Zoomed in images of M96-B5 (top) and M98-B6 (bottom). For each row, there is an HST image right in 1999, shortly after the microlensing events (left), a NIRC2 image from 2016 (center),  and the PSF of the NIRC2 image as determined by \texttt{AIROPA} (right). The sky images are each 2" across, with a large light-blue circle with radius 0.3" centered on each target. All sources found by \texttt{AIROPA} are indicated with smaller red circles. In both cases, no new sources have appeared within 0.3" of the source, which corresponds to the maximum separation of a lens moving with proper motion of ~15 mas/yr.}
	\label{fig:zooms}
\end{figure*}

\subsection{Keck Reduction and Star-finding with AIROPA}\label{sec:starfinder}

Initial reduction of raw data from each epoch of observation was carried out with our custom NIRC2 reduction pipeline \citep{2008ApJ...675.1278S,2009ApJ...690.1463L} and included flat field and dark calibration, sky subtraction and cosmic ray removal. The cleaned exposures were corrected for distortion and achromatic differential atmospheric refraction, shifted to a common coordinate system, and then combined, weighted by Strehl using the IRAF routine, \texttt{Drizzle} \citep{2002PASP..114..144F}, as described in \cite{2010ApJ...725..331Y}.  The final combined image  was restricted to individual frames displaying core FWHM$<$1.25 FWHM$_{min}$, where FWHM$_{min}$ is the minimum FWHM of all frames of the particular target and epoch. 

On each combined map, we used the point-spread function (PSF) fitting routine, \texttt{AIROPA} \citep{witzel2016airopa}, which is based on \texttt{StarFinder}, \citep{2000Msngr.100...23D} to extract a stellar catalog of spatial coordinates and relative brightness. 
\texttt{AIROPA} is used in single-PSF mode, which assumes the PSF is uniform over the field of view. First, the PSFs of a subset of stars (hereafter “PSF stars”) were averaged to extract a mean PSF. The PSF was then cross-correlated with the image and stars were identified as peaks with a correlation above 0.8.  Additionally, we ran \texttt{StarFinder} in 'deblend' mode in order to check for a difference in sensitivity to close pairs of stars. 
There was no difference in the number of stars detected within the central 0\farcs5 of either target (the range within which it would be reasonable to find a resolved lens). As such, \texttt{AIROPA} was run in normal ('non-deblend') mode for the remainder of this work.

The accuracy of our PSF directly impacts the contrast (i.e., our ability to detect a faint lens near the brighter source) as well as astrometric precision. Thus, we applied strict criteria in the selection of the PSF stars. Specifically, stars contributed to the mean PSF derivation if they were bright (typically Kp \textless 18 mag), isolated, and within 4\arcsec of the center of 

\noindent the field. The latter criterion avoids detector edge effects and ensures all PSF stars are close to the target of interest, mitigating errors due to spatial variation of the PSF caused by instrumental aberrations and atmospheric anisoplanatism. 

The resulting starlists produced by \texttt{AIROPA} contain positions and fluxes for each star in detector units of pixels and counts, respectively. Residual images with found sources removed were also created for each target/epoch. Instrumental magnitudes were calibrated to Kp using  J, H, and Ks magnitudes from the VVV Survey DR2 \citep{2017yCat.2348....0M}. 

\subsection{HST Reduction with \texttt{img2xymrduv}} \label{sec:img2xym}
In order to extract photometry from the archival HST data, we used the FORTRAN program \texttt{img2xymrduv}, developed for use with WFPC2 images \citep{anderson2006psfs}. 
The code uses criteria set by the user to find stars in an image before fitting them to a PSF model. The output includes instrumental magnitudes derived from the fluxes, in ADU,  determined by the PSF fits well as distortion corrected centroid positions.

To calibrate the magnitudes produced by  \texttt{img2xymrduv}, we divided the observed flux in ADU by the total exposure time and applied the VEGAMAG system zero points for WFPC2  \citep{gonzaga_biretta__2010}.
However, the zero points are defined for counts measured in a 0\farcs5 radius aperture; thus we applied an additional $\Delta ZP$ to calibrate the  \texttt{img2xymrduv} magnitudes extracted from a smaller aperture. We determined $\Delta ZP$ by performing aperture photometry on the calibrated images with a 0\farcs5 radius aperture, positioned on the centroids output by  \texttt{img2xymrduv}.  A sigma clipped median pixel value in each image was used as the sky value to subtract off the aperture sums, and the resulting counts $N_{aper}$ were then propagated through the equation 
\begin{equation}
m_{aper}=-2.5 \log_{10}(\frac{N_{aper} \cdot g} {t_{exp}})
\label{eq:mag}
\end{equation}
along with the gain $g$ and exposure time $t_{exp}$ to get the magnitude, $m_{aper}$.  The differences between $m_{aper}$ and the \texttt{img2xymrduv} magnitudes for the brightest 10\% of stars in an image was median combined to define $\Delta ZP = m_{aper} - m_{img}$. Finally, the calibrated magnitudes were computed as
\begin{equation}
m = m_{img} + ZP + \Delta ZP
\label{eq:zp}
\end{equation}
where $m_{img}$ is the \texttt{img2xymrduv} determined magnitude and $ZP$ is the zero point pulled from \cite{holtzman1995photometric}.

\subsection{Comparison of Archival and Current Images} \label{sec:comp}
Equipped with starlists from both our late time (2016-2017) Keck images and early time (1999-2003), we can now determine if a new luminous lens was detected. In the case that a lens is a black hole, there should be no source nearby the late time images what wasn't present in the early time images. Searching within a radius of 3" of the source (which corresponds to a maximum proper motion of  15.2 mas/yr for M96-B5 and 16.7 mas/yr for M98-B6), no new sources appear. Visual comparisons of these images for both targets is shown in Figure \ref{fig:zooms}. To understand the limitations of this comparison, however, we must examine our detection sensitivity.

\section{Detection Limits on luminous Lenses}\label{sec:limits}

\subsection{Star Planting}\label{sec:plants}

\begin{figure}
	\centering
	\includegraphics[width=0.5\textwidth]{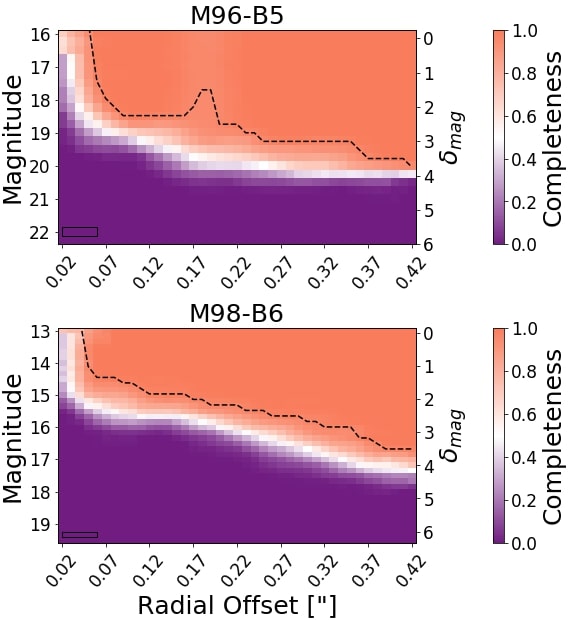}
	\caption{ Completeness curves for MB 96-5 (top) and MB 98-6 (bottom). 95\% completeness is indicated by dashed black lines. By finding where the detection completeness for a giving magnitude goes below 0.95, we determined the magnitude range of a stellar lens undetectable by \texttt{Starfinder}. These magnitude ranges gave us our upper limits to the mass of an undetectable, but luminous, stellar lens.}
	\label{fig:curves}
\end{figure}

To estimate the detection sensitivity of the images, a series of star planting simulations was conducted. 
Individual stars were planted in each image by taking a copy of the point spread function (PSF) that was created using the original image. 
In order to reduce computational time, the main image and background image were cropped from a 1064$\times$1064 pixel array down to 200$\times$200 pixels, and the PSF was cropped down from 200$\times$200 pixels 30$\times$30 pixels.
Normally, the PSF is normalized such that the integral over the entire PSF is 1. 
However, because the cropped PSF (cPSF) was taken from an initially normalized PSF, the integral over the cropped PSF is less than 1. 
The cPSF was then calibrated to a chosen magnitude. Poisson noise was then added to the cPSF. This calibrated and noisy cPSF was then used as an artificial star. The artificial star was then planted in the image, with its center located at pixel position $x_i, y_i$ on the image. A total of 167,445 artificial stars were planted for each event, varying in magnitude from 12.00 to 23.00, with steps of 0.25, and in location from -30 to 30 pixels in both the x- and y-directions around the target. 
The resulting image, with a single planted star, was analyzed with \texttt{AIROPA} in an identical manner to that used on the original image, using the exact same PSF. It was put through \texttt{StarFinder} to see if the planted star was detected. The planted star was considered detected if its position in the new starlist matched the input location to within a single pixel in both
the x- and y-direction and if its magnitude agreed with the input to within 0.5 mag.

A radial completeness curve was created by binning the artificial stars into magnitude bins of 0.25 mag and radial bins of using a sliding window of 39.72 $mas$ (4 $pixels$). 
Within each magnitude-radius bin, the completeness was determined by calculating the percentage of planted stars within the category that were detected. 
A completeness value of 1.0 corresponds to a detection percentage 100\%, and a value of 0.0 corresponds to a detection percentage of 0\%. 
Limiting magnitudes were determined for each radial bin by taking the faintest magnitude with a completeness value of at least 0.95. 
The limiting magnitudes were given the average magnitude of each bin and the average radius of each window (Figure \ref{fig:curves}). No limiting magnitudes were used that exceeded the brightness of the respective source.

\subsection{Isochrones Based on Magnitude}\label{sec:iso}

The limiting magnitudes were converted to stellar masses, as described below, and used to determine the allowed mass range for an undetected, luminous stellar lens.
Synthetic isochrones were generated using the program \texttt{SPISEA} (Hosek et al. in prep). Isochrones were generated for a given age in $log(yr)$, distance in $pc$, and extinction value $A_{k}$, assuming solar metallicity. 
Distances ranged from 1 $kpc$ to 10 $kpc$, with steps of 1 $kpc$. 
The extinction values, which are dependent on the Galactic coordinates and distance of the source, were determined using the Argonaut Skymaps \citep{2015ApJ...810...25G} \citep{2018arXiv180103555G}, which output an $E(B-V)$ value. That $E(B-V)$ was then used to calculate an $A_{V}$ value by assuming $R_{V}=3.1$. 
Then, the extinction value $A_{k}$ was calculated using the extinction law from \citep{2011ApJ...737..103S}. 
As a proxy for a "zero-age main-sequence", we adopted an age of 100 $Myr$ to obtain a nearly fully main sequence while also excluding pre-main sequence stars. 

Using the generated isochrones, the outputted mass values were taken for any isochrone entries that matched a limiting magnitude to within 0.2 mag. 
Thus, for each limiting magnitude, corresponding to a radial angular separation from the source, a mass-distance relationship was determined. 
We note that these mass limits are extracted under the assumption that the lens is a main-sequence star. However, pre-main-sequence or post-main-sequence stars are more luminous for a given mass; thus our use of a main-sequence mass-luminosity relationship is conservative. Finally, each radial bin was converted into a proper motion bin using the respective time elapsed since closest approach and the distance value inputted into the synthetic isochrone. With this, our completeness in magnitude/radial seperation space has been projected to mass/lens distance/relative proper motion space. This new parameterization allows our completeness to be related to the microlensing events, the light curves of which are described by physical quantities including lens mass $M$, lens distance $D_L$, and relative proper motion $\mu_{rel}$.  


\section{Light Curve Fitting and Analysis}\label{sec:lightcurves}

\subsection{Fitting Routine and Results}\label{sec:fits}

\begin{figure*}
	\centering
	\includegraphics[width=\textwidth]{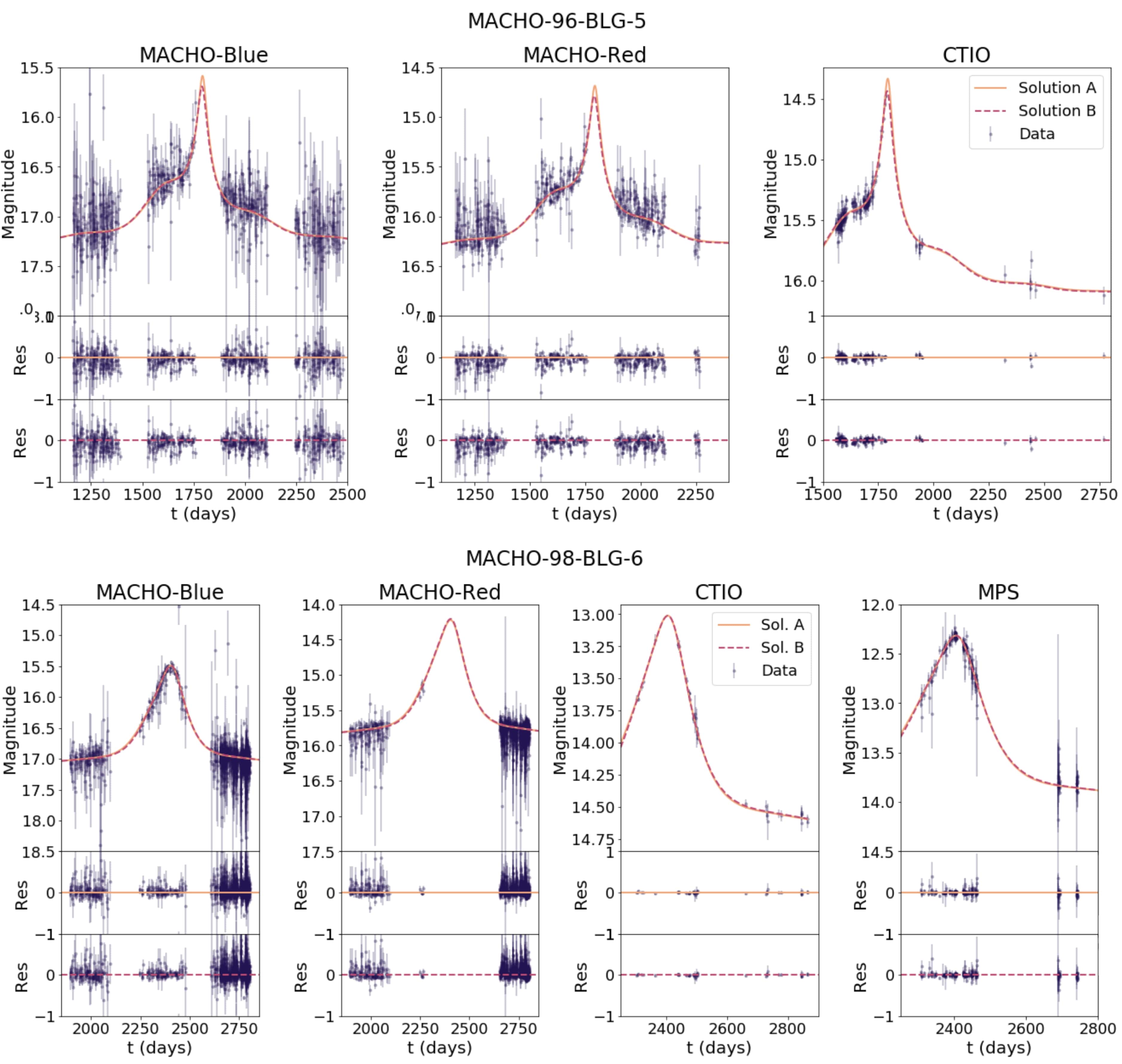}
	\caption{ Fitted models of the light curves for M96-B5 (top) and M98-B6 (bottom). In each case, the light curve data is shown in dark purple points, the best fit model with negative $u_{\textrm{0}}$ in solid orange, and the best fit model with positive $u_{\textrm{0}}$ in dashed magenta. The residuals (data - model) for each light curve are shown immediately below in matching colors and style. Note that the visualization of the data has been truncated to only show rising/falling of the light curve, and not baseline (or approximately baseline) measurements. The date $t$ is reported in JD - 2448623.5.}
	\label{fig:fits}
\end{figure*}
Microlensing models were fit to the light curves described in Section \ref{sec:macho} by applying the
publicly available software package \texttt{pyLIMA}, using the differential evolution fitting method \citep{bachelet2017pylima}. Each event was fit to a standard point-source point-lens (PSPL) model with parallax. The number of free parameters in the model depends on the number of light curves of different filters fit. For any single-lens parallax event, there are initially five parameters: the time of closest approach between the source and lens ($t_\textrm{0}$); the impact parameter between the source and the lens in units of the Einstein radius ($u_{\textrm{0}}$), the characteristic timescale, or 'Einstein crossing time' ($t_{\textrm{E}}$); and the North and East components of the microlensing parallax vector ($\pi_{\textrm{E,N}}$ and $\pi_{\textrm{E,E}}$). For each filter, there are two additional parameters: the assumed-to-be-static flux of the source in the absence of lensing, $f_s$ and the blend flux $f_b$, which is the flux from companions to the lens, companions to the source, and/or ambient interloping stars that incidentally fall within the seeing disc of the microlensing target (the 'target' being the superimposed lens and source) during the event. The values obtained by the fits for this PSPL model for both events are presented in Table \ref{tab:fits}. Corner plots showing posteriors for all solutions are shown in Figure 12-15.

Additionally, after each initial fit was conducted, a second fit was run with the additional constraint that $u_{\textrm{0}}$ only be allowed to have values with the opposite sign of the initial $u_{\textrm{0}}$ results, in order to find degenerate solutions. For example, if $u_{\textrm{0}}$ was first found to equal -0.5, in the second fit, $u_{\textrm{0}}$ was restricted to positive values.

The two flux parameters can be recast to a more intuitive pair of parameters: un-lensed baseline magnitude of the source ($m_s$) and blend-source-flux-fraction ($b_{sff}$), which is the ratio of the flux from the source alone to the total flux in the aperture of the images from which the light curves were derived. The transformation is made with the following equations:

\begin{equation}
m_s = 27.4-2.5 \log_{10}(f_s)
\label{eq:fs}
\end{equation}
\begin{equation}
b_{sff} = \frac{f_s}{f_s+f_b}
\label{eq:bsff}
\end{equation}

The resulting $m_s$ and $b_{sff}$ values for a subset of the data (that will be used in the following section) are presented in Table \ref{tab:flens}.

\begin{deluxetable*}{lcccc}
	\tabletypesize{\footnotesize}
	\tablecolumns{11} 
	\tablewidth{0pt}
	\tablecaption{Light Curve Model Fit Parameters}
	\tablehead{
		Parameter & M96-B5 (A) & M96-B5 (B) & M98-B6 (A) & M98-B6 (B)	}
	\startdata
	$t_{\textrm{0}}$ [days] & 1766.2$^{+1.2}_{-1.3}$ & 1773.10$^{+0.88}_{-0.93}$ & 2413.7$^{+1.4}_{-1.5}$ & 2407.4$^{+1.5}_{-1.6}$ \\ 
	$u_{\textrm{0}}$ [$\theta_{\textrm{E}}$] & -0.0296$^{+0.0049}_{-0.0052}$ & 0.0164$^{+0.0042}_{-0.0037}$ & -0.174$^{+0.019}_{-0.022}$ & 0.175$^{+0.023}_{-0.023}$ \\ 
	$t_{\textrm{E}}$ [days] & 537$^{+92}_{-68}$ & 628$^{+100}_{-74}$ & 367$^{+41}_{-37}$ & 346$^{+45}_{-35}$ \\ 
	$\pi_{EN}$ & 0.0397$^{+0.0074}_{-0.0070}$ & -0.0284$^{+0.0050}_{-0.0051}$ & -0.087$^{+0.023}_{-0.021}$ & 0.0294$^{+0.023}_{-0.024}$ \\ 
	$\pi_{EE}$ & 0.076$^{+0.012}_{-0.012}$ & 0.0535$^{+0.0073}_{-0.0077}$ & 0.076$^{+0.012}_{-0.013}$ & 0.0873$^{+0.0066}_{-0.0076}$ \\ 
	$f_s$ (MACHO-B) & 1900$^{+330}_{-320}$ & 1550$^{+240}_{-240}$ & 9600$^{+1400}_{-1200}$ & 9300$^{+1500}_{-1500}$ \\ 
	$f_b$ (MACHO-B) & 9110$^{+250}_{-270}$ & 9400$^{+170}_{-190}$ & 3900$^{+1100}_{-1400}$ & 4300$^{+1400}_{-1500}$ \\ 
	$f_s$ (MACHO-R) & 4320$^{+750}_{-710}$ & 3510$^{+540}_{-540}$ & 31400$^{+5500}_{-4700}$ & 31000$^{+5500}_{-5000}$ \\ 
	$f_b$ (MACHO-R) & 21820$^{+540}_{-590}$ & 22470$^{+390}_{-420}$ & 10300$^{+4500}_{-5300}$ & 11000$^{+4800}_{-5300}$ \\ 
	$f_s$ (CTIO) & 6100$^{+1100}_{-990}$ & 5000$^{+770}_{-770}$ & 96000$^{+14000}_{-12000}$ & 93000$^{+15000}_{-14000}$ \\ 
	$f_b$ (CTIO) & 26300$^{+1100}_{-1200}$ & 27200$^{+1000}_{-1000}$ & 31000$^{+11000}_{-13000}$ & 35000$^{+14000}_{-15000}$ \\ 
	$f_s$ (MPS) &  - & - & 183000$^{+27000}_{-23000}$ & 177000$^{+29000}_{-28000}$ \\ 
	$f_b$ (MPS) & - & - & 56000$^{+21000}_{-25000}$ & 63000$^{+27000}_{-28000}$ \\ 
	$\chi^2_{\nu}$ & 1.57 & 1.56 & 1.17 & 1.18 \\
	d.o.f. & 2290 & 2290 & 2302 & 2302 \\
	\enddata
	\tablenotetext{}{Results of fitting MACHO light curves to PSPL models. Each event has two solutions, with negative and positive $u_{\textrm{0}}$ due to the degeneracy in this parameter, henceforth referred to 'Solution A' and 'Solution B', respectively.  Note that the date $t_{\textrm{0}}$ is JD-2448623.5.}
	\label{tab:fits}
\end{deluxetable*}

The PSPL with parallax model well describes the M98-B6 light curve, fitting with a reduced chi-squared of 1.17 and 1.18 for negative and positive $u_{\textrm{0}}$ values, respectively. As is visible in the bottom panel of Figure \ref{fig:fits}, this event appears to be a smooth PSPL curve with minor perturbation due to parallax. The fit for M96-B5 is slightly poorer, with reduced chi-squared of 1.57 and 1.56; though both models appear to fit the data well, the lack of coverage for times of peak amplification introduces some uncertainty, yielding more variation in solutions than for M98-B6, which has full peak coverage in two filters.

To potentially better fit M96-B5, we attempted fitting to a point-source-binary-lens (PSBL) model, which has three additional parameters: the log of the mass ratio between the two lenses $\log{q}$; the log of the projected binary separation in units of the Einstein radius $\log{s}$; and the position angle between the binary axis and source trajectory $\alpha$. However, because the PSBL model did not significantly increase the quality of the fit ($\chi^2_{\nu}$=1.62), we will only consider the PSPL results for the remainder of this work.

\subsection{Constraining $f_{lens}$}\label{sec:flens}

In cases of a luminous lens or additional light sources within the aperture of the source star, the resultant microlensing light curve has substantial alterations to it. The additional, unlensed flux remains constant while the flux of the source varies, resulting in a smaller apparent maximum magnification, as well as what appears to be a shorter $t_{\textrm{E}}$. This information is captured in $b_{sff}$, which we can rewrite as as 
\begin{equation}
b_{sff}=\frac{f_s}{f_s+f_b}=\frac{f_s}{f_s+f_l+f_n}
\label{eq:bsff_split}
\end{equation}
where $f_l$ is the flux from the lens and $f_n$ is the flux from neighboring stars. $b_{sff}$ is fit as a free parameter when modeling the photometric light curve.

To constrain $f_{l}$, we can combine $b_{sff}$ with two relevant quantities derived from the WFPC2 images. The first quantity is the instrumental magnitude of the target from \texttt{img2xymrduv} as described in Section \ref{sec:img2xym}. Paired with image exposure time, this magnitude is converted to a count flux which we interpret as $f_s+f_l$ (which we will call $f_{targ}$). As the HST images were taken only 2-3 years after the peak of each event, we assume here that the source and lens would have not had adequate time to separate appreciably. Second, by summing the total flux within the 1.2” radius circle centered on the source, imitating the observations from the original MACHO data set, we obtain $f_s+f_l+f_n$ (or $f_{aper}$).  We can write the relationships between these quantities: 

\begin{equation}
\bigg( \frac{f_s + f_l}{f_s + f_l + f_n} - \frac{f_s}{f_s + f_l + f_n} \bigg) (f_s + f_l + f_n)=f_l
\label{eq:flens1}
\end{equation}
\begin{equation}
f_l = \bigg(\frac{f_{targ}}{f_{aper}} - b_{sff}\bigg)f_{aper}
\label{eq:flens2}
\end{equation}

thereby extracting the lens flux.  This quantity is then converted back into a WFPC2 instrumental magnitude and finally calibrated into Vega magnitude. The corresponding source magnitude is derived similarly. Note that this process is only possible for the components of the light curve data in the MACHO-Blue and MACHO-Red filters, as their passbands align reasonably well to that of the WFPC2 F555W and F675W filters (shown in the appendix).

\begin{figure*}[]
	\centering
	\includegraphics[width=\textwidth]{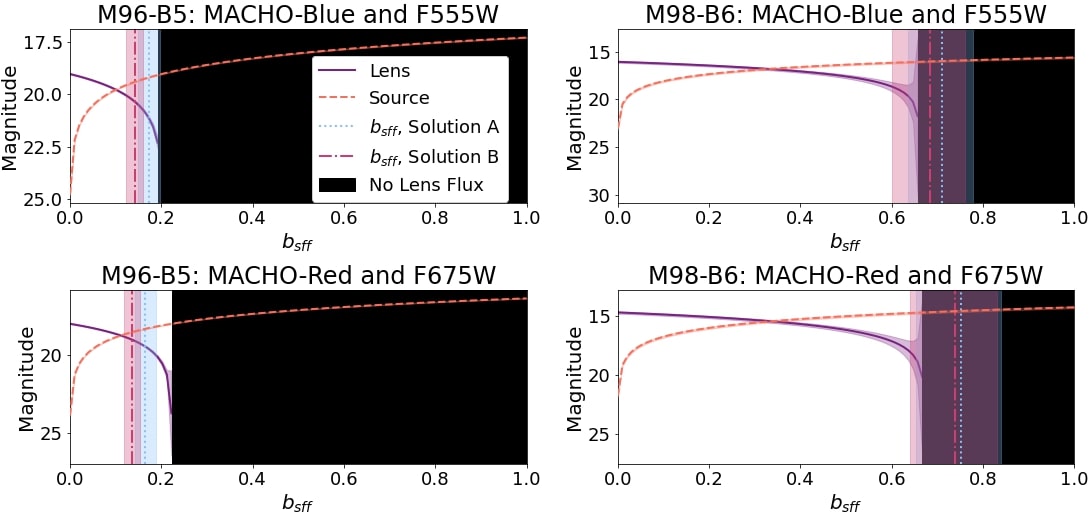}
	\caption{For each target and filter combination, the solid purple curve shows the magnitude of the lens for all possible values of $b_{sff}$. The dashed orange curve shows the corresponding source magnitude. The vertical lines show the values obtained for $b_{sff}$ from Table \ref{tab:fits}, in each case with Solution A ($-u_{\textrm{0}}$) values light blue and dotted, and Solution B ($+u_{\textrm{0}}$) magenta and dot-dashed. The black regions designate the values of $b_{sff}$ that would result in zero lens flux. The intersections between the lens curve and vertical lines are this work's estimates of the lens' magnitudes. Where the value of $b_{sff}$ falls within a region on zero lens flux but has an error bar that extends into the region of non-zero lens flux, the intersection of the lower $b_{sff}$ boundary and the solid purple curve are taken as lower limits on the lens magnitude.}
	\label{fig:flens}
\end{figure*}

This analysis was done on all HST frames in the epoch 2000-06-11 from Table \ref{tab:hst}. Though there are data of M96-B5 in 1999, they only exists in the F555W filter, and was thus excluded in favor of a slightly later epoch for the purpose of combining the MACHO-Blue/F555W results with the MACHO-Red/F675W results in a later section. Figure \ref{fig:flens} shows a summary of these results, where source and lens magnitude are plotted as a function of $b_{sff}$ from 0 (no source flux) to 1 (no non-source flux). As seen in this figure, the result is highly sensitive to $b_{sff}$. The source and lens magnitudes derived from this analysis are presented in Table \ref{tab:flens}.

In the case of this work’s value of $b_{sff}$, both solutions for M96-B5 indicate a lens with non-zero flux. For M98-B6, we see that both solutions' $b_{sff}$ indicates that there is no lens flux, which we would expect in the case of a BH lens. However, the 1-sigma error bar on $b_{sff}$ does extend into the region of non-zero lens flux. As such, in Table \ref{tab:flens} we report the lens magnitudes at the lower boundary of $b_{sff}$ regions as limits on the lens magnitude of this event.


\begin{deluxetable*}{lcccccc}
    \tabletypesize{\footnotesize}
    \tablewidth{0pt}
    \tablecaption{Lens Magnitude Constraints}
    \tablehead{Event & Filter Combination & $m_{targ}$ & Solution & $b_{sff}$  & $m_{source}$ & $m_{lens}$}
    \startdata
	M96-B5 & MACHO-B/F555W &  19.027$\pm$ 0.027 & A & 0.173$^{+0.025}_{-0.024}$ & 19.20$^{+0.15}_{-0.16}$ & 21.11$^{+0.89}_{-0.93}$ \\
	               &                                &                                & B & 0.141$^{+0.019}_{-0.019}$  & 19.42$^{+0.15}_{-0.15}$ & 20.32$^{+0.33}_{-0.34}$ \\
	               & MACHO-R/F675W & 17.969$\pm$0.015  & A & 0.165$^{+0.024}_{-0.023}$ & 18.30$^{+0.15}_{-0.16}$ & 19.42$^{+0.43}_{-0.45}$ \\
	               &                                  &    & B                              & 0.135$^{+0.018}_{-0.018}$ & 18.52$^{+0.15}_{-0.15}$ & 18.97$^{+0.22}_{-0.23}$ \\
	M98-B6 & MACHO-B/F555W  & 16.038$\pm$0.044 & A & 0.710$^{+0.067}_{-0.075}$ & 15.96$^{+0.15}_{-0.13}$ & $>$18.45 \\
	               &                                  &     & B                             & 0.683$^{+0.078}_{-0.082}$ & 16.00$^{+0.16}_{-0.15}$ & $>$18.16 \\
	               & MACHO-R/F675W  & 14.677$\pm$0.044 & A& 0.752$^{+0.087}_{-0.10}$ & 14.553$^{+0.17}_{-0.15}$ & $>$18.38 \\
	               &                                  &      & B                            & 0.739$^{+0.091}_{-0.099}$ & 14.573$^{+0.17}_{-0.16}$ & $>$18.44 \\
    \enddata
    \tablenotetext{}{For a given event, light curve solution and filter combination, we present the results of using the blend fraction $b_{sff}$ to separate the target magnitude $m_{targ}$ into the source and lens components, $m_{source}$ and $m_{lens}$, respectively.}
    \label{tab:flens}
\end{deluxetable*}

\section{Source Analysis} \label{sec:sources} 

\subsection{Astrometric Determination of Source Proper Motion}\label{sec:ast}
Before the alignment necessary for computing proper motions for the source of M96-B5, the WFPC2 starlists described in Section \ref{sec:img2xym} were consolidated into single starlists for each epoch and filter.  This was done using follow up programs to \texttt{img2xymrduv}, \texttt{xym2mat} and \texttt{xym2bar}, which match and average combine sources in a set of starlists (Anderson et al. 2008). By combining multiple observations at a given epoch, these collated starlists include uncertainties in centroid positions, which are ultimately needed for establishing the precision of the final proper motion determination.

Though the source star for M96-B5 is not in Gaia EDR3, all stars in the catalog within a 36$\arcsec$ box centered on the source were used as an absolute reference frame to which all epochs of data were matched.  This included both epochs of NIRC2 data and all epochs of WFPC2 data in the F555W and F814W filters. The starlists from these data sets were matched and transformed into the coordinate system of the EDR3 starlist with \texttt{FlyStar}, a package that performs matching and astrometric transformations on starlists. The resulting relative positions are reported in the appendix.

\begin{figure}
	\centering
	\includegraphics[width=0.45\textwidth]{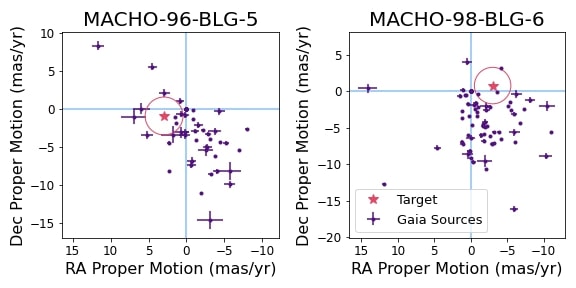}
	\caption{ Vector point diagrams of each target (magenta), each shown with all Gaia sources within a 36" box centered on the respective target (purple). Proper motions are in the Gaia reference frame. A magenta circle centered on each target indicates the region of proper motion space in which an object would have a proper motion relative to the target of 2.5 mas/yr. An object that is within this circle and also positioned within 0\farcs3 of the target would not be resolvable from the source in our 2016/2017 images.}
	\label{fig:pms}
\end{figure}

Once the cross-epoch alignment was complete, the positions of the source at each epoch were fit to a linear, constant proper motion model, yielding a proper motion of -2.955$\pm$0.022 mas/yr in the East direction and -0.926$\pm$0.022 mas/yr in the North direction, with a reduced chi squared of 1.38. The proper motion of this target, along with the proper motions of all Gaia EDR3 sources within a 36" box centered on the target, are shown in the vector point diagram in Figure \ref{fig:pms}.

Unlike M96-B5, the source star of M98-B6 matches a object included in Gaia EDR3. The position, proper motion, and parallax of the source in Gaia are given in Table \ref{tab:gaia}. There are several checks we use to confirm that this object is indeed the same as our microlensing source. First, we consider the ruwe (renormalized unit weight error) parameter, for which values below 1.4 suggest an unblended source with robust astrometric measurments \citep{fabricius2020gaia}. As this source has a ruwe parameter of 1.29, it is unlikely that these are two distinct but spatially-unresolvable stars. Additionally, the ruwe parameter justifies our use of the measured parallax in estimating the source distance (which is 6.7$^{+11.9}_{-2.6}$ kpc). 

To further confirm that the Gaia object is indeed the source star and not an unrelated star at a different distance, the procedure described for determining the proper motion for M96-B5 above was repeated for M98-B6 in order to compare the results to the proper motion reported in Gaia. The result was a proper motion of -2.82$\pm$0.35 mas/yr in the East direction and 0.47$\pm$0.36 mas/yr in the North direction. Within uncertainties, this is consistent with the Gaia proper motions of -2.98$\pm$0.11 mas/yr East and 0.812$\pm$0.075 mas/yr North, further supporting the claim that these two sources are one and the same. The proper motion of this source in the context of its field is also shown in Figure \ref{fig:pms}.

\subsection{Photometric Exploration of Source and Lens}\label{sec:phot}
As shown in Figure \ref{fig:curves}, our ability to directly observe a luminous lens becomes difficult as the radial offset from the source in the image plane decreases. If the relative proper motion between a source and lens is small, there may not be sufficient angular separation between the two objects to resolve them in our NIRC2 images. To explore a scenario in which objects appear blended even after ~20 years, we use the WFPC2 photometry from the 2000-06-11 (for M96-B5) and 2000-06-23 (for M98-B6) epochs to look for indications of additional unresolved objects in/near the source. 
Though this concept was initially intended for the purpose of covering the low relative proper motion case, some back-of-the-envelope calculations show that it should actually hold regardless of proper motion: high lens mass (BH) microlensing events will produce Einstein radii of a few miliarcseconds, which we can use along with the light curve fit’s $u_{\textrm{0}}$ values (see Table \ref{tab:fits}), which is in units of the Einstein radius, to show that at time of closest approach, the lens and source in both events should not have been separated more than $\sim$1 mas. The times between this closest approach and the observations are no more than 3 years, which, along with the aperture radius of 0\farcs5 used for the photometry, means that the relative lens source proper motion would have to be $\sim$100 mas/yr in order for the lens to to have moved outside of the aperture. We can see that for both events, the source proper motion is only a few mas/yr, meaning the majority of the required ~100 mas/yr would have to be made up by the lens. Thus, we can reasonably assume that at the time of the WFPC2 observations, the lens is still within the aperture observing the source.

We will call the hypothetically unresolved lens and source object the ‘target’, for which the photometric information in four filters was given in Table \ref{tab:hst}.  In the case of M96-B6, we also have disentangled photometric data for a lens and source in two filters (F555W and F675W), which were presented in Table \ref{tab:flens}. For M98-B6, the disentangled photometry consists of source magnitudes and lower limits on the lens magnitude.

In order to identify objects that our photometric data could represent, we again use \texttt{SPISEA}, the stellar population generating python package discussed in Section \ref{sec:iso}. \texttt{SPISEA} allows not only for the choice of many parameters for the populations it creates, but also for photometry from particular instruments and filters to be simulated. For the purposes of matching to our data, we simulated photometry for HST’s WFPC2 instrument in the filters F439W, F555W, F675W and F814W. 

To create a sample of stars to search in, we generated isochrones at a range of cluster distances (1 kpc to 18.5 kpc, in increments of 0.5 kpc, the upper limit set by the maximum distance allowed by M98-B6’s EDR3 parallax measurement) and ages (log age in years spanned 8 to 10, in increments of 0.5). Metallicity was held constant at solar metallically, and default evolution and atmospheric models were chosen.  A Cardelli extinction law with Rv=3.1 was applied along with extinction values pulled from the previously cited Argonaut Skymaps. This resulted in a library of 50 isochrones with ~200 stars each. The code does not generate compact objects or brown dwarfs. 

\begin{figure*}[]
	\centering,
	\includegraphics[width=\textwidth]{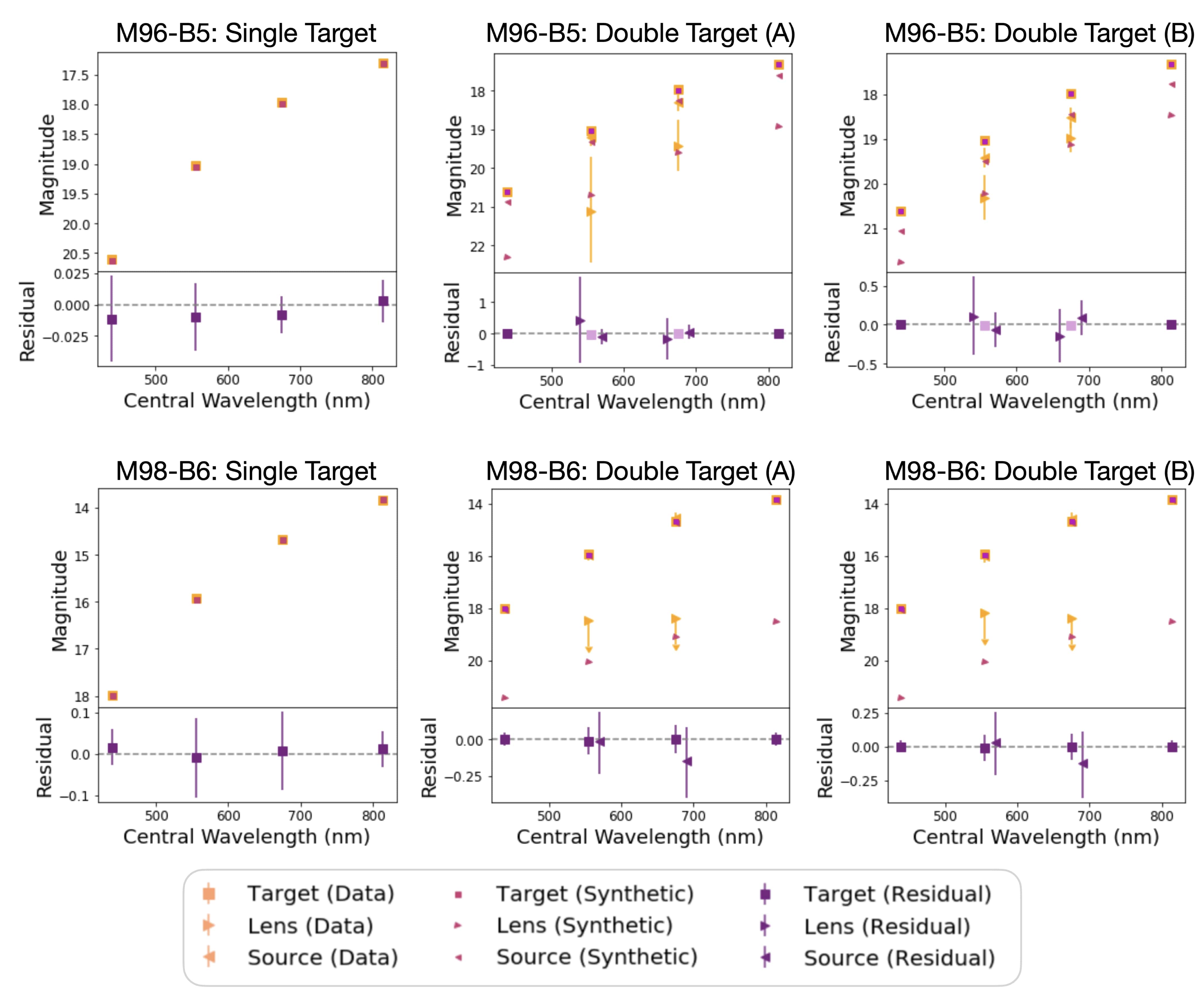}
	\caption{Comparisons of broadband photometry of the event targets to best-matched synthetic data. For each event (M96-B5 in the top row and M98-B6 in the bottom row), three potential scenarios are considered. In the left-most panel, the target is assumed to be only the background source (meaning the lens is either non-luminous or is spatially resolved from the source). In the middle and right-most panels, the target is assumed to be an unresolved source and luminous lens. The disentanglement of the lens and source in these cases is derived from the two solutions to each light curve fit, Solution A in the center and Solution B on the right. In each figure, the total target data is represented in orange squares, and if a separate lens and source are shown, they are represented in orange triangles directed to the right and left, respectively. The synthetic data matches are symbolically represented the same, but in magenta with smaller markers. In the residual panels at the bottom of each plot, if lens and source points are present, they are slightly offset from their central wavelength to improve figure visibility. Additionally, only the dark purple points were used to calculate $\chi^2$, as the light purple points are redundant information, but useful for visualization.}
	\label{fig:cmds}
\end{figure*}

For each event, we explored three scenarios. In the first, we made the assumption that the WFPC2 photometric data from the 2000 epoch (see Table \ref{tab:hst}) showed only a single object and thus looked for the closest single-object match.  This was done by identifying the minimum chi-squared between the photometric data and each synthetic photometry object. Though our light curve fits for M96-B5 indicated a flux contribution from both a lens and source, Figure \ref{fig:flens} shows how sensitive this result is to $b_{sff}$, and though our result allowed for non-zero lens flux, previous works that have fit the same light curve have obtained blending parameters that, in the analysis in Section \ref{sec:flens} would yield zero lens flux. For example, the fit from Bennet 2002 presented two solutions that would put $b_{sff}$ fully in the black region of Figure \ref{fig:flens}). We note that though our fits yielded lower chi-squared than those of Bennet 2002, there is inherent ambiguity in this event due to the lack of coverage of the peak magnification. As such, we explore this possibility along with the scenarios more supported by this work.

The second scenario and third scenarios we explore for each event is the case of a single luminous source and single luminous lens, with magnitudes (or magnitude limits, for the lens of M98-B6) dependent on the two light curve solutions. These disentangled lens and source magnitudes come from Section \ref{sec:flens}, in which we derived separate lens and source photometry in the filters F555W and F675W. The least squares search for this scenario operated slightly differently than the single object case. For M96-B5, the data points matched were the F555W and F675W photometry of one object for the source and one object for the lens, in addition to the F439W and F814W photometry for the summed source and lens combination (as there was not adequate data to separate the lens and source contribution in these two filters). Because M98-B6 has lens magnitude limits rather than specific values, the search was conducted by first finding all objects that fell above the lens limits, then summing each of these with any objects at a greater distance. For each of these combinations, chi-squared was calculated with the two source points (still F555W and F675W) and all four of the combined target points, instead of any lens points. In either case, all six points used were weighted equally in the search.

For each of the four scenarios above, a closest matched source (or sources and lens) was identified. The physical parameters of the matched sources, as well as a chi squared per point are presented in Table \ref{tab:phot}. Comparisons of the data to the synthetic photometry for each scenario are illustrated in Figure \ref{fig:cmds}). The isochrones from which each match object was pulled are illustrated in Figure \ref{cmds_app}, in the appendix.

\begin{deluxetable*}{llcccccccc}[h]
    \tabletypesize{\footnotesize}
    \tablewidth{0pt}
    \tablecaption{Best Synthetic Photometry Matches}
    \tablehead{Event & Scenario & Component &  $\log{\textrm{A}_{c}}$ (y) & D$_c$ (kpc) & T$_{*}$ (K) & M$_{*}$/M$_{\odot}$ & L$_{*}$/L$_{\odot}$ &  R$_{*}$/R$_{\odot}$ & $\chi^2/$pts
    } 
     \startdata  
     M96-B5 & Single Target          & Source & 9.5 & 8.5 & 4960 & 1.41 & 10.76 & 4.44 & 0.144 \\
     \hline
            & Double Target (A)      & Source & 9.5 & 8 & 5043 & 1.41 & 7.39 & 3.56 & 0.079 \\
            &                        & Lens   & 9 & 2.5 & 4584 & 0.73 & 0.17 & 0.66 & \\
    \hline
            & Double Target (B)      & Source  & 9.5 & 10 & 4980 & 1.41 & 9.76 & 4.19 & 0.082 \\
            &                        & Lens    & 8 & 2  & 4457 & 0.70 & 0.14 & 0.63 & \\
    \hline
     M98-B6 & Single Target          & Source & 8.5 & 7.5 & 4572 & 3.19 & 297.30 & 27.46 & 0.057 \\
     \hline
            & Double Target (A)      & Source & 8 & 16.5 & 4633 & 5.11 & 1466 & 59.36 & 0.056 \\
            &                        & Lens   & 10 & 3.5 & 5352 & 0.84 & 0.54 & 0.85 & \\
    \hline
            & Double Target (B)      & Source  &  8 & 16.5 & 4633 & 5.11 & 1466 & 59.36 & 0.045 \\
            &                        & Lens   &  10 & 3.5 & 5352 & 0.84 & 0.54 & 0.86 & 
    \enddata 
    \tablenotetext{}{Closest matches from synthetic isochrones to target photometry. For each scenario, the log age and distance to the cluster(s) in which the closest match  was found is given, in addition to the effective temperature, mass, luminosity, and radius of the matching object. As a metric of fit, chi-squared per point is reported. For single target scenarios, chi-squared is calculated with four data points, while double target scenarios are calculated with six.}
    \label{tab:phot}
\end{deluxetable*}

\section{RESULTS} \label{sec:results}

\begin{figure*}[]
	\centering,
	\includegraphics[width=\textwidth]{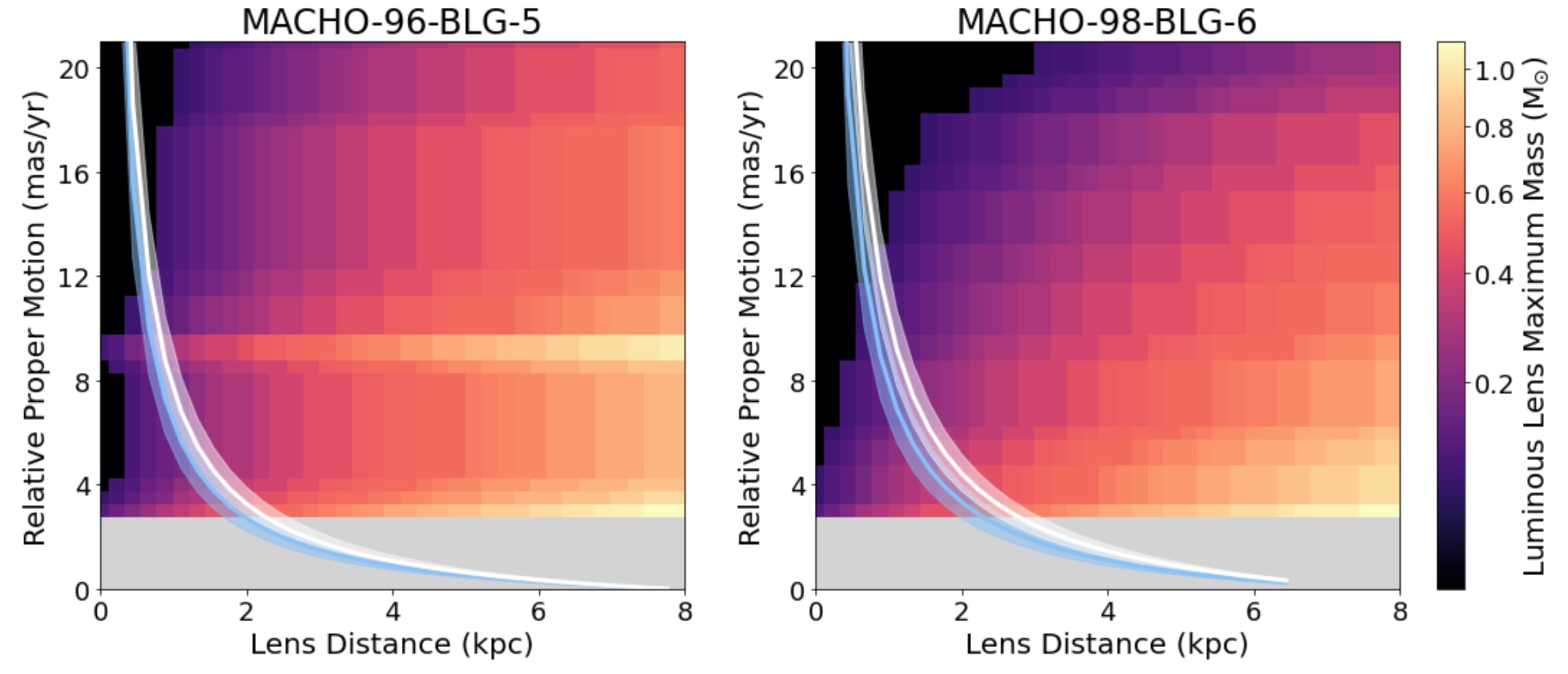}
	\caption{For each event, the colorbar indicates the maximum mass of a stellar lens that could have been undetected by our process of artificial star planting and retrieval in NIRC2 images described in Section \ref{sec:limits} as a function of lens distance and relative proper motion, with the black area indicating parameter space for which there were no stellar matches found. The lens distance/relative proper motion relationship derived from each solution of the light curve fit is shown in cyan and white lines, with transparent surrounding regions designating plus or minus one sigma to that relationship, with Solution A in cyan and Solution B in white. The gray region represents the possibility of relative proper motions low enough ($<2.5$ mas/yr) such that the lens and source are not resolvable. The only stellar lenses allowed by our model fits and not identified in our NIRC2 images must be below the masses indicated by the colorbar within the cyan and white bands.}
	\label{fig:results}
\end{figure*}

\begin{figure*}[]
	\centering,
	\includegraphics[width=\textwidth]{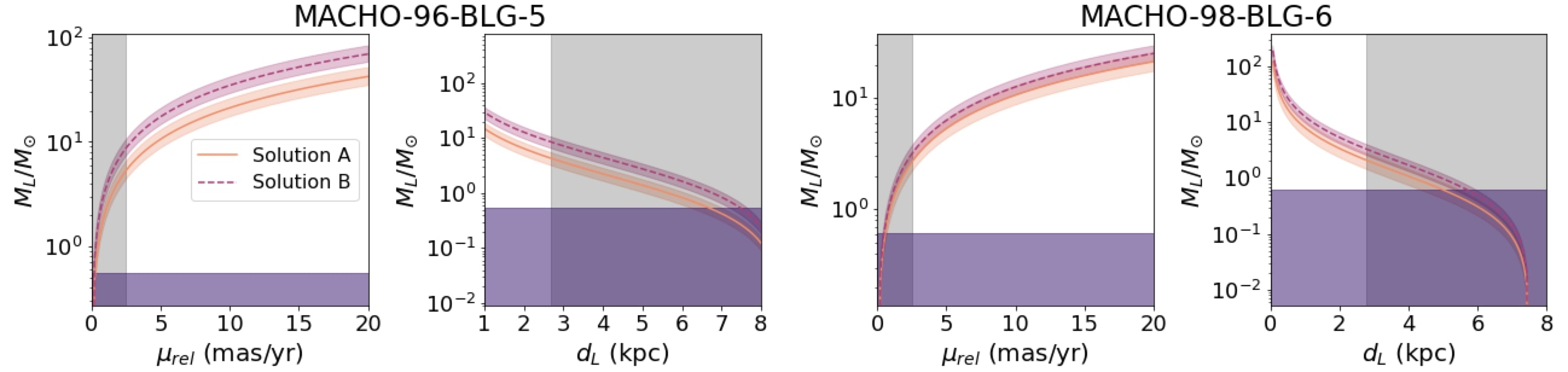}
	\caption{Possible lens masses as a function relative lens-source proper motion (first and third panels from the left) and as a function of lens distance (second and fourth panels from the left) for both events. In each case, the solid orange and dashed magenta lines represent the x-axis variables' relationship to lens mass as determined by Solution A and B of the light curve fit results, respectively, with shading designating plus or minus 1-sigma to that relationship. The purple regions identify all possible lens masses at which a stellar lens would have been undetected by our star finder. The gray region represents the range of relative proper motions low enough and corresponding lens distance limits such that the lens and source are not resolvable. In the regions in which we are to resolve a lens and source, undetectable masses fall below the allowed masses from the light curves in every case. Note that the curves assume a source distance; in each case, the source distance assumed is that of the best-fit photometric match to a single target, given in Table \ref{tab:phot}.}
	\label{fig:results2}
\end{figure*}

To encapsulate the results of the three analysis sections of this paper (4-6), we will begin by looking at Figure \ref{fig:results}, which summarizes the findings of Sections \ref{sec:limits} and \ref{sec:fits} and illustrates the need for Sections \ref{sec:flens} and \ref{sec:sources}. 

For each event, the colorbar indicates the maximum mass of a stellar lens that could have been undetected by our process of artificial star planting and retrieval in NIRC2 images described in Section \ref{sec:limits}. This mass limit is presented as a function of lens distance and relative lens-source proper motion, two microlensing parameters that cannot be constrained by a light curve alone. The black area indicates parameter space for which there were no stellar matches found. Note that despite being cast into parameters that are meaningful for microlensing events, the completeness data itself is derived solely from our 2016/17 images of the sources, and not the microlensing events. 

The photometric microlensing light curve data is incorporated in the cyan and white curves (for Solution A and B, respectively), which indicate combinations of lens distance and relative lens-source proper motions allowed by the fits in Section \ref{sec:fits}. The surrounding shaded cyan and white regions designate plus or minus one sigma to those relationships. The curves assume a source distance: 8.5 kpc for M96-B5 and 7.5 kpc for M96-B5 (each chosen based on the the best-matched synthetic object to a single source target, reported in Table \ref{tab:phot}). The only stellar lenses allowed by our model fits and not identified in our NIRC2 images must be below the masses indicated by colors within these shaded cyan and white bands.

Combining these two pieces of analysis of two different data sets, we find that the maximum possible mass of a stellar lens in each event is 0.53 $M_{\odot}$ and 0.55 $M_{\odot}$ for Solutions A and B of M96-B5, and 0.55 $M_{\odot}$ and 0.61 for M98-B6. We note that these mass limits are derived from simulating main-sequence stars, which are less luminous for a given mass than pre-main-sequence or post-main-sequence stars (meaning these types of stars would all be detectable as well). However, we also note that this does not account for other luminous objects, such as brown dwarfs or non-BH compact objects.

These masses, however, have little to do with the lens masses allowed by the light curve data. By combining Equations 1-3, we can derive a microlensing event's lens mass as a function of the fit parameters $t_{\textrm{E}}$ and $\pi_{\textrm{E}}$, and the relative proper motion $\mu_{rel}$:

\begin{equation}
\label{eq:murelMass}
    M=\frac{t_{\textrm{E}}\mu_{rel}}{\kappa \pi_{\textrm{E}}}
\end{equation}

 where $\kappa=\frac{4G}{1\:\textrm{AU}\:c^2}$. Further, by assuming a source distance as described above, we can recast this relationship as lens mass as a function of lens distance using Equations 2-3:

\begin{equation}
\label{eq:MassDist}
    M=\frac{1\:\textrm{AU}\:(d_L^{-1}-d_S^{-1})}{\kappa\pi_{\textrm{E}}^2}
\end{equation}

Both relationships for each event and solution are illustrated in Figure \ref{fig:results2}, in which, similar to Figure \ref{fig:results}, each panel has a gray region which indicates relative proper motions too low (or corresponding lens distances too high) to resolve a lens and source (<2.5 mas/yr). To compare to Figure \ref{fig:results}, each panel of Figure \ref{fig:results2} also includes a purple region along the bottom shading all masses below 1.2 $M_{\odot}$, the highest possible undetectable stellar mass based on our NIRC2 image completeness. In all cases outside the gray regions, the range of lens masses allowed by the light curve fit is well above the range of stellar masses undetectable in our images, eliminating the possibility of a stellar lens in this parameter space. At the boundary of this region, we find that the lens masses indicated by a relative proper motion of 2.5 mas/yr are 5.30$^{+1.14}_{-0.96}$ M$_{\odot}$ and 8.72$^{+1.70}_{-1.46}$ M$_{\odot}$ for solutions A and B of M96-B5, and 2.67$^{+0.53}_{-0.49}$ M$_{\odot}$ and 3.17$^{+0.52}_{-0.48}$ M$_{\odot}$ for solutions A and B of M98-B6. As all of these are above even a conservative upper limit of neutron star mass of 2.16 M$_{\odot}$ \citep{rezzolla2018using}, this eliminates the possibility of non-BH compact objects such as neutron stars and white dwarfs, as well as low mass, luminous objects such as brown dwarfs or free floating planets.

The only remaining exception is the possibility of relative proper motions low enough such that the lens and source would not have had adequate time to separate appreciably, in which case a potentially luminous lens would not be detectable in our NIRC2 images, regardless of mass. This possibility is represented by the gray regions at the bottom of each panel in Figure \ref{fig:results} and along the sides of each panel in Figure \ref{fig:results2}. To address this blindspot, Sections \ref{sec:flens} and \ref{sec:sources} defined and explored possible scenarios with photometric and astrometric analysis of WFPC2 images shortly following the events in 2000, the results of which are in Table \ref{tab:phot}. We discuss the implications and limitations of these results in the following section.

\section{DISCUSSION} \label{sec:discussion}

\subsection{Implications of Light Curve Fit Results} \label{sec:dis1}

From an examination of the light curve fit parameters alone (in Table \ref{tab:fits}), we have some clues to the nature of the lens. We first note the exceptionally long $t_{\textrm{E}}$, particularly in the case of M96-B5. Because $t_{\textrm{E}}$ scales with the square-root of lens mass, these long timescales (as opposed to the days-long events caused by planets and months-long events caused by stars and lower-mass compact objects) suggest very high masses. Indeed, it would be difficult for lens stars below $\sim$0.6 $M_{\odot}$ (the masses defined as ‘allowed’ in the previous section) to result in microlensing events with timescales of hundreds of days. 
	
\begin{figure}
	\centering
	\includegraphics[width=0.45\textwidth]{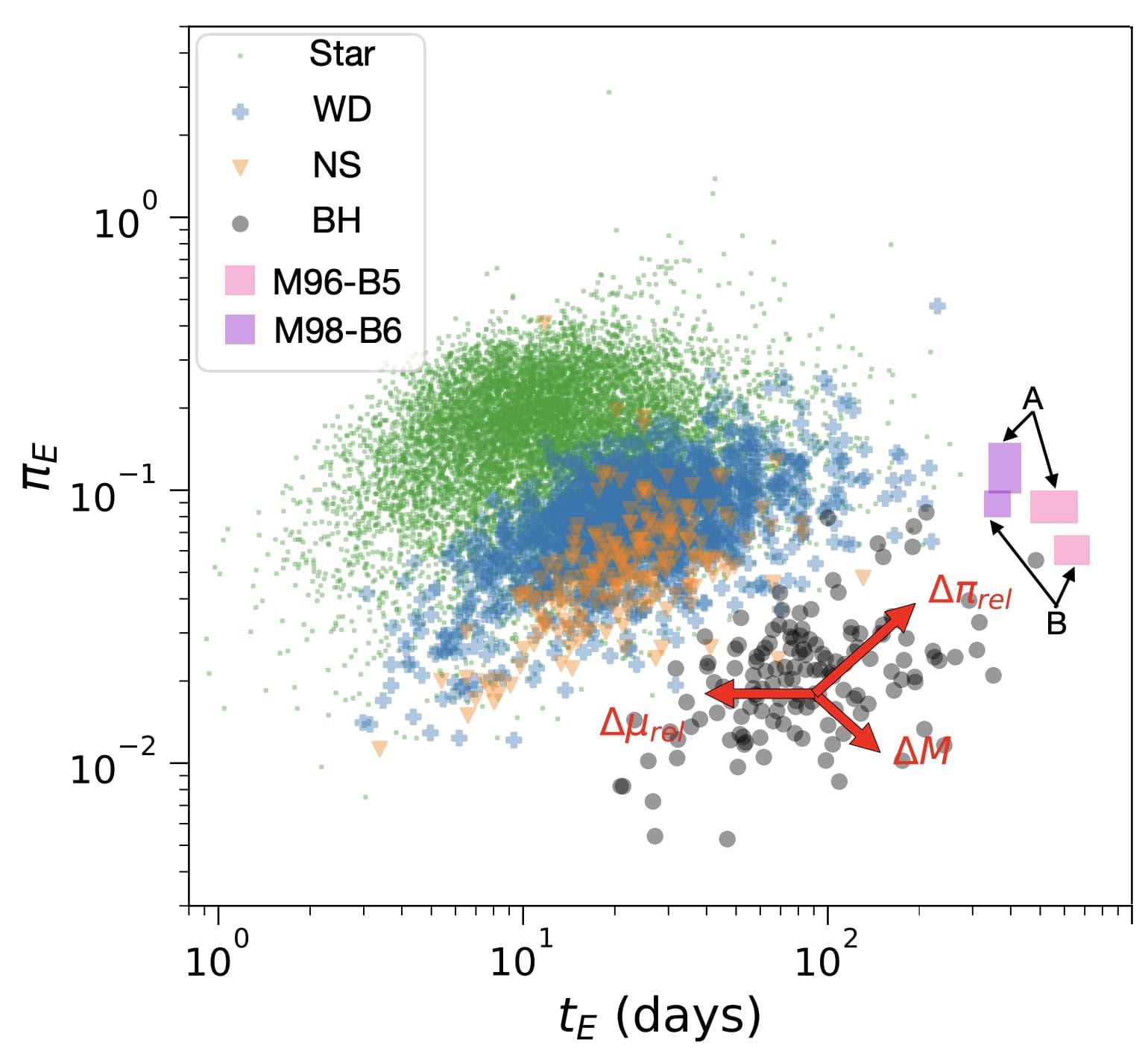}
	\caption{The main panel from Figure 13 in \cite{lam2020popsycle}, showing the distributions of $t_{\textrm{E}}$ nd $\pi_E$ by lens type for a simulated microlensing survey. The one sigma ranges of these parameters from the results of this work's light curve fits are shown in magenta and purple boxes, for M96-B5 and M98-B6, respectively. The two different solutions for each event are marked with 'A' and 'B' arrows in black.}
	\label{fig:lam}
\end{figure}

If we pair our events’ $t_{\textrm{E}}$ values with another output of our fits, $\pi_E$ (the ‘microlensing parallax’, defined in Equation \ref{eq:piE}), we can put them in the context of the simulations constructed in \cite{lam2020popsycle}, which emulate the results of microlensing surveys while examining the distribution of lens and source objects. One figure from this work, (replicated here as Figure \ref{fig:lam}), shows the distribution of star, white dwarf, neutron star, and black hole lenses as a function of $t_{\textrm{E}}$ and $\pi_E$. The magenta and purple boxes have been added on the figure to show where our events (M96-B5 and M98-B6, respectively) fall in this distribution. This work reports that if searching for BH lenses in microlensing events, the BH detection rate will be ~85\% among events with $t_{\textrm{E}}>120$ days and $\pi_{\textrm{E}}<0.08$. Solutions B of M96-B5 meets this criteria fully, while solution B of M96-B5 and both solutions of M98-B6 meet the $t_{\textrm{E}}$ cutoff, but have $\pi_{\textrm{E}}$ values going up to 0.11.

In the context of the \cite{lam2020popsycle} result, the values of $\pi_\textrm{E}$ and $t_{\textrm{E}}$ for both events suggest a similarity to microlensing events simulated with compact objects lenses, though a stellar lens is not impossible. However, considering that the detection limits for this event indicate that a stellar lens would only be possible with a relatively low ($\lessapprox 0.6M_{\odot}$) mass, the fact that the event falls much closer to the high mass (lower right) corner of the distribution than the low mass (top right) corner of the distribution suggests that the lens is not a low mass main sequence star. Though this isn't an explicit elimination of a luminous, low mass lens in either case, it strongly suggests that the low mass stars left as potential lenses in our analysis are not likely lens candidates.
	
The possibility of low mass luminous sources are further diminished by the results shown in Figure \ref{fig:results2}. For all relative lens-source proper motions above 2.5 mas/yr, the lens masses allowed by the light curve fit at any given proper motion or lens distance are well above the range of stellar masses undetectable in our NIRC2 images, effectively ruling out all stellar lenses in this region of parameter space. Further, we rule out any lens object in this parameter space with mass below 5.30$^{+1.14}_{-0.96}$ M$_{\odot}$ and 2.67$^{+0.53}_{-0.49}$ M$_{\odot}$ for M96-B5 and M98-B6 respectively, eliminating the possibility of a low-mass, non-stellar lens, including neutron stars, white dwarfs, brown dwarfs, and free-floating planets. In effect, there is no possibility for a non-BH lens in either event for relative proper motions above 2.5 mas/yr. We can reverse this calculation to limit the proper motions even further by recognizing that no luminous lens above 2.16 M$_{\odot}$ is possible (as the results in Table \ref{tab:phot} show that an the maximum possible masses for a stellar lens don’t exceed 1 M$_{\odot}$, and all non-stellar objects besides BHs would fall below that limit). Calculating the maximum proper motion allowed at this mass limit with Equation \ref{eq:murelMass}, we find that, using the more conservative result for each event, non-BH lenses are only possible at relative proper motions below 1.32 mas/yr and 2.48 mas/yr for M96-B5 and M98-B6, respectively. The remainder of this discussion evaluates the likelihood of a BH lens in the parameter space below these proper motion limits, in which the lens and source would be unresolved in our images. 

\subsection{Blend Flux Variability in Lens Flux Analysis} \label{sec:dis2}

Light curve modeling of M98-B6 shows  no indication of a luminous lens at all proper motions (though the 1-sigma error bar extends slightly into the parameter space of non-zero lens flux). This is not the case for M96-B5, which, for both solutions appear to have a non-zero amount of lens flux in our results. We note, however, that the result of this analysis is highly sensitive to the value of $b_{sff}$ (which is calculated from the fit parameters $f_s$ and $f_b$). In using this parameter as the basis of this analysis, we must consider the potential for degeneracies between this parameter, $t_{\textrm{E}}$, and $u_{\textrm{0}}$. An event with long $t_{\textrm{E}}$ but low values for $u_{\textrm{0}}$ and $b_{sff}$ will be degenerate with a short $t_{\textrm{E}}$ event with higher $u_{\textrm{0}}$ and $b_{sff}$. 
	
We can look at previous fits in the literature to see this degeneracy manifest, such as the results for fitting the M96-B5 light curve to a PSPL model with parallax in \cite{2002Ben}. This work reported three solutions for this event, one of which is similar to our best fit solution (long $t_{\textrm{E}}$, small $u_{\textrm{0}}$ and $b_{sff}$), while the other two illustrated the opposite result (relatively short $t_{\textrm{E}}$, with larger values of $u_{\textrm{0}}$ and $b_{sff}$). The values yielded in the latter two solutions for $b_{sff}$  were 0.3 and 0.33 in MACHO-Blue and 0.28 and 0.31 in MACHO-Red. If applied to the lens flux analysis here, these results would put the lens for M96-B5 squarely in the region of no lens flux, which, paired with the long $t_{\textrm{E}}$ and small $\pi_E$ in the context of the \cite{lam2020popsycle} cited result above, strongly suggest a BH lens. To address this limitation of fitting (as noted previously, the lack of coverage of the peak magnification of this event introduces an inherent ambiguity), we explored a non-luminous lens scenario in addition to the two luminous lens scenarios for this target in Section \ref{sec:phot}, which will be further discussed below.
	
Similarly, by looking at the result obtained for a PSPL with parallax fit of M98-B6 in \cite{2005ApJ...633..914P}, we find three physical solutions for this event (excluding the fourth solution, which has a negative, non-physical blending parameter). All three yielded lower $t_{\textrm{E}}$ values than our reported $\sim$350 days, but $b_{sff}$ values larger than ours (.76, .85, and 1). This only supports our Section \ref{sec:flens} result that for this event, there is no measurable contribution of flux from a lens.

\subsection{Interpreting Closest Matches in Target Photometry} \label{sec:dis3}

Because of the degeneracy involving $b_{sff}$ and the impact it has on our results, we took a conservative approach in using the analysis of Section \ref{sec:flens} to inform the photometric examination of the source in Section \ref{sec:phot} and considered both non-luminous and luminous lens scenarios for both events. Note that ‘target’ here refers to the unresolved object which contains at least light from the source star, and at most light from the source star and unidentified lens object.

For M96-B5, the scenario in which we consider the target as a single source (the scenario that the Bennett 2002 solution’s high $b_{sff}$ values would suggest), we find that the best match 1.41 M$_{\odot}$ G9 sub-giant at 8.5 kpc. This distance corresponds to a star in the galaxy’s bulge, which is typically assumed for microlensing events observed in that direction. The two scenarios with a source and a luminous lens have similar objects matched to the source (both 1.41 M$_{\odot}$ G9 sub-dwarfs, at 8 kpc and a 10 kpc for solutions A and B), with low mass main-sequence stars matched to the lens (a K2 at 2.5 kpc and a K3 at 2 kpc, both with masses $\sim$0.7 M$_{\odot}$). Though the quality of the match for the lens+source scenarios appears to be better than the single source scenario (chi-squared values of 0.079 and 0.082 for the double objects versus 0.144 for the single object), if considered in the context of Equation \ref{eq:MassDist}, the combinations of lens distance, source distance, and lens mass of either of the double object matches is not allowed by corresponding light curve fit. As such, we report finding no combined source and luminous lens photometric matches that are consistent with all of our data for this event.

The same can be said about our results for M98-B6 (for which our analysis in Section \ref{sec:flens} had already pointed to zero-lens flux, with a small possibility of a luminous lens). Though in all three scenarios, the matched source yields an object with temperature and distance consistent with the EDR3 effective temperature and parallax measurements, all lenses within our magnitude limits yielded masses too low to be allowed with the corresponding source and lens distances. Further, the single source is well described consistent with the classification of a previous work \citep{Soto:2007} which used optical spectral fitting to classify this source as a G5 sub giant.

\subsection{Remaining Possibilities and Future Observations} \label{sec:dis4}

In discussing what remains, we first examine the limitations of the synthetic photometry matching. As discussed previously, the potential matches are pulled from synthetic isochrones that include main-sequence and post-main-sequence stars. This does not account for the possibility of non-stellar lenses, including neutron stars, white dwarfs, and brown dwarfs. Further, objects such as these may be missed by our lens flux analysis, as it was limited by observations in only optical wavelengths. 

For M96-B5, this parameter space is narrowed down by several X-ray observations of the event, which found the X-ray emission to be consistent with a BH lens \citep{maeda2005search, 2006ApJ...651.1092N}. This leaves white dwarfs, brown dwarfs, and free floating planets as possible remaining lenses for this event. By Equation \ref{eq:murelMass}, a white dwarf lens at the Chandrasekar limit of 1.4 M$_{\odot}$ would necessitate a relative proper motion of 0.66$^{+0.15}_{-0.12}$ mas/yr for solution A and 
0.40$^{+0.08}_{-0.07}$ mas/yr for solution B. Without X-ray observations, our upper limit on relative proper motion is set by a conservative maximum neutron star mass of 2.16 M$_{\odot}$ \citep{rezzolla2018using}, which yields relative proper motions of 2.02$^{+0.46}_{-0.34}$ mas/yr and
1.71$^{+0.31}_{-0.24}$ mas/yr. By take the upper limit of the more conservative solution for each event, we conclude that only a BH lens is possible for relative proper motions above 0.81 mas/yr for M96-B5 and 2.48 mas/yr for M98-B6.

The results presented here indicate that both M96-B5 and M98-B6 remain extremely good candidates for BH lenses; though we cannot make a solid confirmation, multiple methods of examination have yielded no compelling alternatives for either event. There remains, however, additional observations and analysis that could illuminate the true nature of the lenses further.  Primarily, we are concerned with eliminating the possibility of non-BH compact object or brown dwarf lenses unresolved from their source stars. This could be done by spectroscopic examination of the source and searching for a second object in its spectrum, similar to this work’s analysis of broad band photometry. The wavelength of observation would depend on the potential lens: near-IR for brown dwarfs, optical or UV for white dwarfs, and UV or X-ray for neutron stars. Measurement of the source spectrum would yield a secondary benefit for M96-B5: an estimation of the source distance from spectral typing (and perhaps a better constrained distance than that from the Gaia parallax for M98-B6). This would allow for better constraints on our results, which use source distances based on photometric fitting. These straightforward observations would conclusively determine if anything remains in the narrow region of parameter space in which a BH is not the only possibility. 

\section{CONCLUSIONS} \label{sec:conclusion}
Through the analysis of high-resolution images and light curve data, we have eliminated the possibility of a non-BH lens for relative lens-source proper motions above 0.81 mas/yr for M96-B5 and 2.48 mas/yr for M98-B6. To address the potential for an unresolved, luminous lens, our comparison of source images to synthetic photometry indicate that for both events, a single stellar source is a better fit than a source blended with a stellar lens, eliminating the possibility of a stellar lens at any proper motion. We discussed the unlikely, but physically possible scenario of brown dwarf or non-BH compact object lenses with extremely low relative proper motion, and described how this remaining possibility can be constrained by future observations and analyzed with the methods developed in this work.


\section{ACKNOWLEDGEMENTS} \label{sec:acknowledgements}
The majority of this work was completed at U.C. Berkeley, which sits on the territory of xučyun, the ancestral and unceded land of the Chochenyo speaking Ohlone people, the successors of the historic and sovereign Verona Band of Alameda County. We acknowledge that we have benefited and continue to benefit from the use and occupation of this land. In this acknowledgement, we recognize the importance of taking actions to support the rematriation of indigenous land, and pledge to take and continue action in support of American Indian and Indigenous peoples.

The data presented herein were obtained at the W. M. Keck Observatory, which is operated as a scientific partnership among the California Institute of Technology, the University of California and the National Aeronautics and Space Administration. The Observatory was made possible by the generous financial support of the W. M. Keck Foundation.

The authors wish to recognize and acknowledge the very significant cultural role and reverence that the summit of Maunakea has always had within the indigenous Hawaiian community.  We are most fortunate to have the opportunity to conduct observations from this mountain.

This work has made use of data from the European Space Agency (ESA) mission Gaia (https://www.cosmos.esa.int/gaia), processed by the Gaia Data Processing and Analysis Consortium (https://www.cosmos.esa.int/web/gaia/dpac/consortium). Funding for the DPAC has been provided by national institutions, in particular the institutions participating in the Gaia Multilateral Agreement.

The first author thanks Etienne Bachelet for his assistance with \texttt{pyLIMA}, as well as Calen Henderson, Ellie Schwab Abrahams, Dan Weisz, and the anonymous referee for helpful discussions and comments.
\clearpage

\section{Appendix}

\begin{deluxetable}{lcccc}[h]
\tablecaption{M96-B5 Astrometric Measurments}
\tabletypesize{\footnotesize}
\tablewidth{0pt}
\label{tb:ast96}
\tablehead{Epoch & Instrument & Filter & RA Offset (mas) & Dec Offset (mas) 
}
\startdata
1999.452 & WFPC2 & F555W & 53.1 $\pm$ 2.3 & 14.8 $\pm$ 0.9 \\
1999.452 & WFPC2 & F814W & 53.1 $\pm$ 2.0 & 16.3 $\pm$ 1.2 \\
2000.444 & WFPC2 & F555W & 49.3 $\pm$ 1.4 & 15.5 $\pm$ 2.3 \\
2001.419 & WFPC2 & F555W & 46.3 $\pm$ 1.4 & 16.4 $\pm$ 4.0 \\
2001.419 & WFPC2 & F814W & 47.5 $\pm$ 0.5 & 14.5 $\pm$ 0.3 \\
2001.747 & WFPC2 & F555W & 45.4 $\pm$ 3.9 & 13.8 $\pm$ 3.4 \\
2001.750 & WFPC2 & F814W & 47.5 $\pm$ 0.9 & 14.0 $\pm$ 0.9 \\
2002.394 & WFPC2 & F555W & 43.2 $\pm$ 2.8 & 16.9 $\pm$ 3.1 \\
2002.394 & WFPC2 & F814W & 43.5 $\pm$ 1.9 & 15.7 $\pm$ 1.3 \\
2002.750 & WFPC2 & F555W & 43.1 $\pm$ 1.8 & 11.8 $\pm$ 1.8 \\
2002.750 & WFPC2 & F814W & 43.2 $\pm$ 1.0 & 13.2 $\pm$ 1.0 \\
2003.400 & WFPC2 & F555W & 39.9 $\pm$ 1.1 & 15.6 $\pm$ 1.3 \\
2003.400 & WFPC2 & F814W & 40.3 $\pm$ 0.6 & 14.6 $\pm$ 0.7 \\
2016.535 & NIRC2 & Kp & 2.07 $\pm$ 0.10 & 0.49 $\pm$ 0.20 \\
2017.387 & NIRC2 & Kp & 0.00 $\pm$ 0.13 & 0.00 $\pm$ 0.21 \\
\enddata 
\tablenotetext{}{Position of M96-B5 from 1999 to 2017. RA and Dec offset are measure in miliarcseconds, where the final epoch is placed at 0,0.}
\end{deluxetable}

\begin{deluxetable}{lcccc}[h]
\tablecaption{M98-B6 Astrometric Measurments}
\tabletypesize{\footnotesize}
\tablewidth{0pt}
\tablecaption{M98-B6 Astrometry}
\label{tb:ast98}
\tablehead{Epoch & Instrument & Filter & RA Offset (mas) & Dec Offset (mas) 
}
\startdata
2000.476 & WFPC2 & F555W & 49.3 $\pm$ 2.3 & 2.5 $\pm$ 6.1 \\
2000.476 & WFPC2 & F675W & 36.4 $\pm$ 4.1 & -5.9 $\pm$ 2.0 \\
2000.476 & WFPC2 & F814W & 41.7 $\pm$ 1.6 & -5.2 $\pm$ 2.5 \\
2015.500 & Gaia & G & 0.0 $\pm$ 0.1 & 0.0 $\pm$ 0.1 \\
2016.535 & NIRC2 & Kp & -1.9 $\pm$ 0.5 & 1.6 $\pm$ 0.6 \\
2017.436 & NIRC2 & Kp & -4.7 $\pm$ 0.5 & 3.1 $\pm$ 0.8 \\
\enddata 
\tablenotetext{}{Position of M98-B6 from 2000 to 2017. RA and Dec offset are measure in miliarcseconds, where the Gaia epoch (2015.5) is placed at 0,0.}
\end{deluxetable}

\begin{figure}[h]
	\centering
	\label{filters}
	\includegraphics[width=0.45\textwidth]{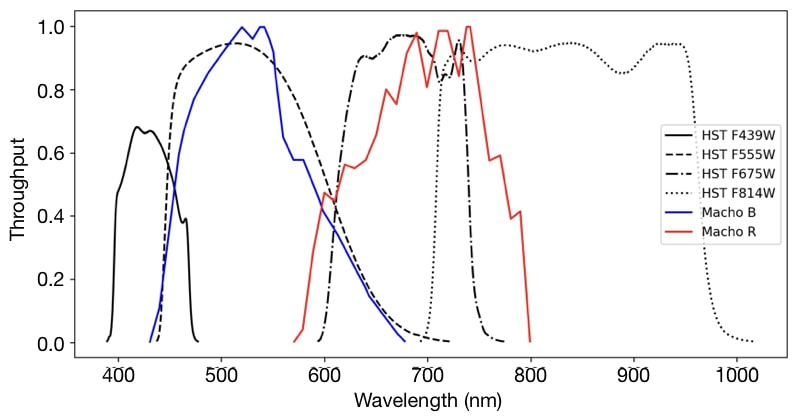}
	\caption{Transmission curves for the two light curve filters (MACHO-Blue and MACHO-Red) compared to the WFPC2 filters used in the analysis of $f_l$. We compare MACHO-Blue to F555W and MACHO-Red to F675W.}
\end{figure}

\begin{figure*}[h!]
	\centering
	\label{corner1}
	\includegraphics[width=\textwidth]{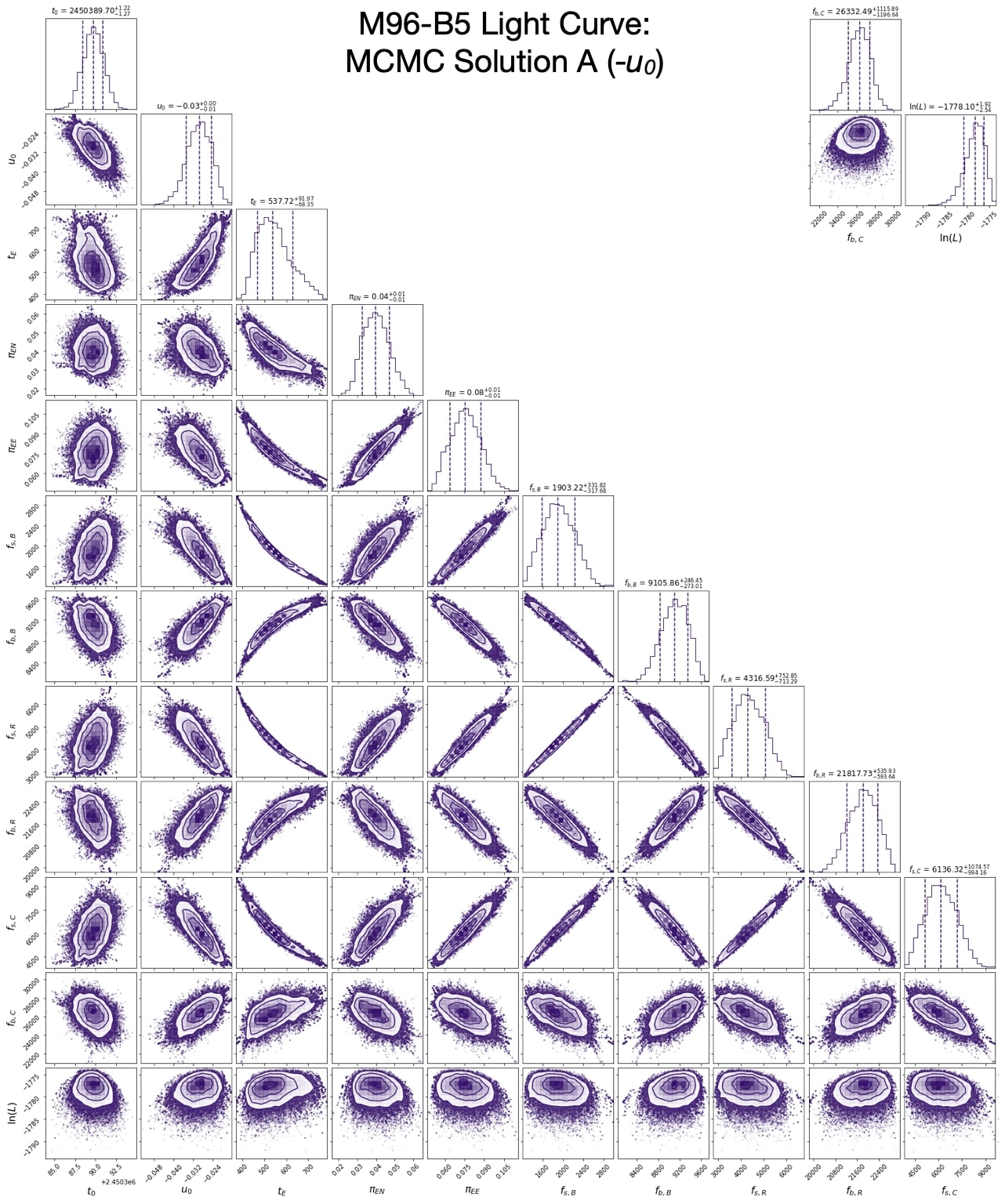}
	\caption{Posterior distributions from one of two MCMC solutions for the light curve of M96-B5, here with negative $u_{\textrm{0}}$. In addition to the 11 parameters defined for Table \ref{tab:fits}, there is an additional row showing the log-likelihood $\ln{L}$, defined here as $-0.5\chi^2$. Note that the rightmost two columns have been re-positioned to fit the figure space.}
\end{figure*}

\begin{figure*}[h!]
	\centering
	\label{corner1}
	\includegraphics[width=\textwidth]{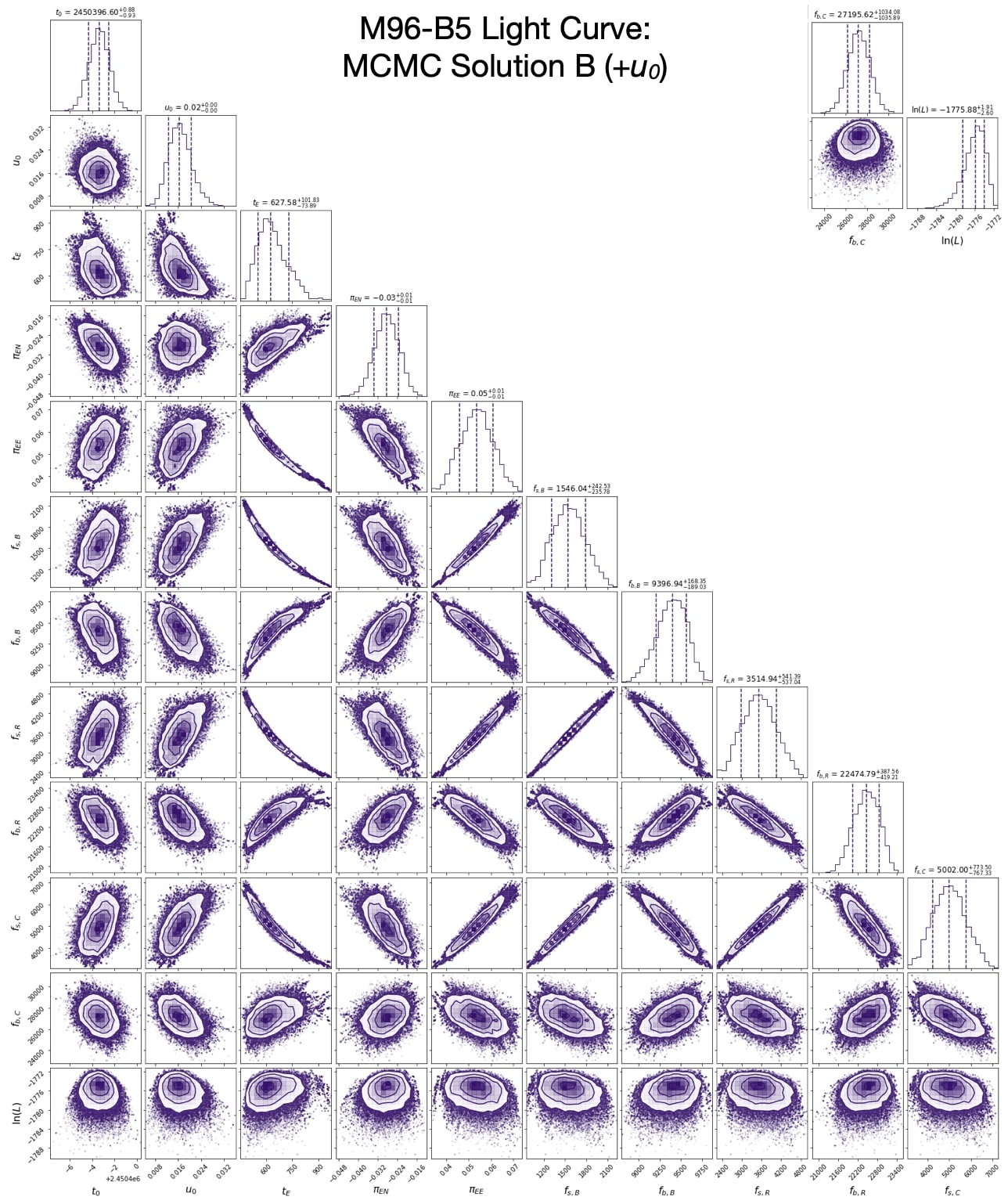}
	\caption{Posterior distributions from one of two MCMC solutions for the light curve of M96-B5, here with positive $u_{\textrm{0}}$. In addition to the 11 parameters defined for Table \ref{tab:fits}, there is an additional row showing the log-likelihood $\ln{L}$, defined here as $-0.5\chi^2$. Note that the rightmost two columns have been re-positioned to fit the figure space.}
\end{figure*}

\begin{figure*}[h!]
	\centering
	\label{corner1}
	\includegraphics[width=\textwidth]{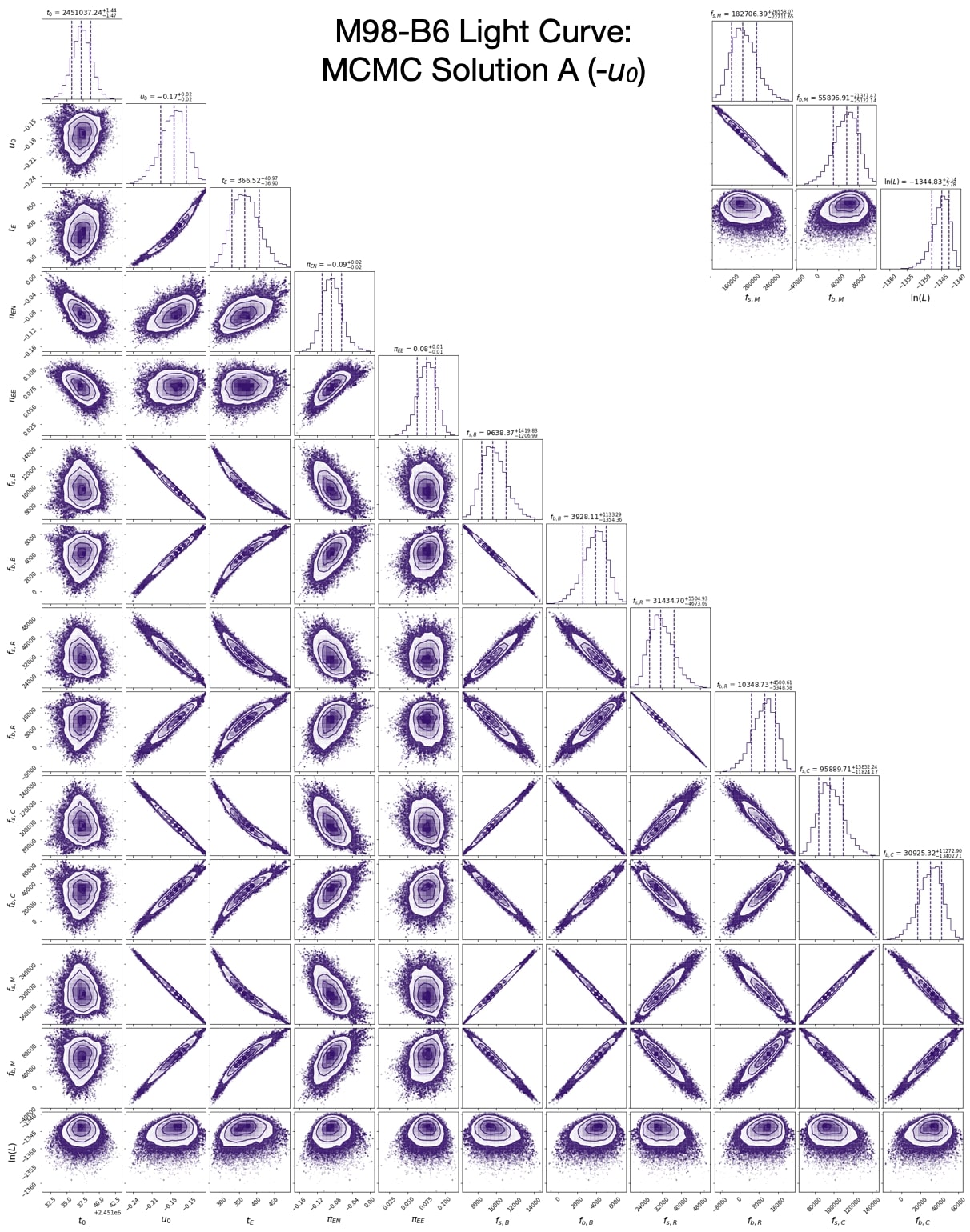}
	\caption{Posterior distributions from one of two MCMC solutions for the light curve of M98-B6, here with negative $u_{\textrm{0}}$. In addition to the 13 parameters defined for Table \ref{tab:fits}, there is an additional row showing the log-likelihood $\ln{L}$, defined here as $-0.5\chi^2$. Note that the rightmost three columns have been re-positioned to fit the figure space.}
\end{figure*}

\begin{figure*}[h!]
	\centering
	\label{corner1}
	\includegraphics[width=\textwidth]{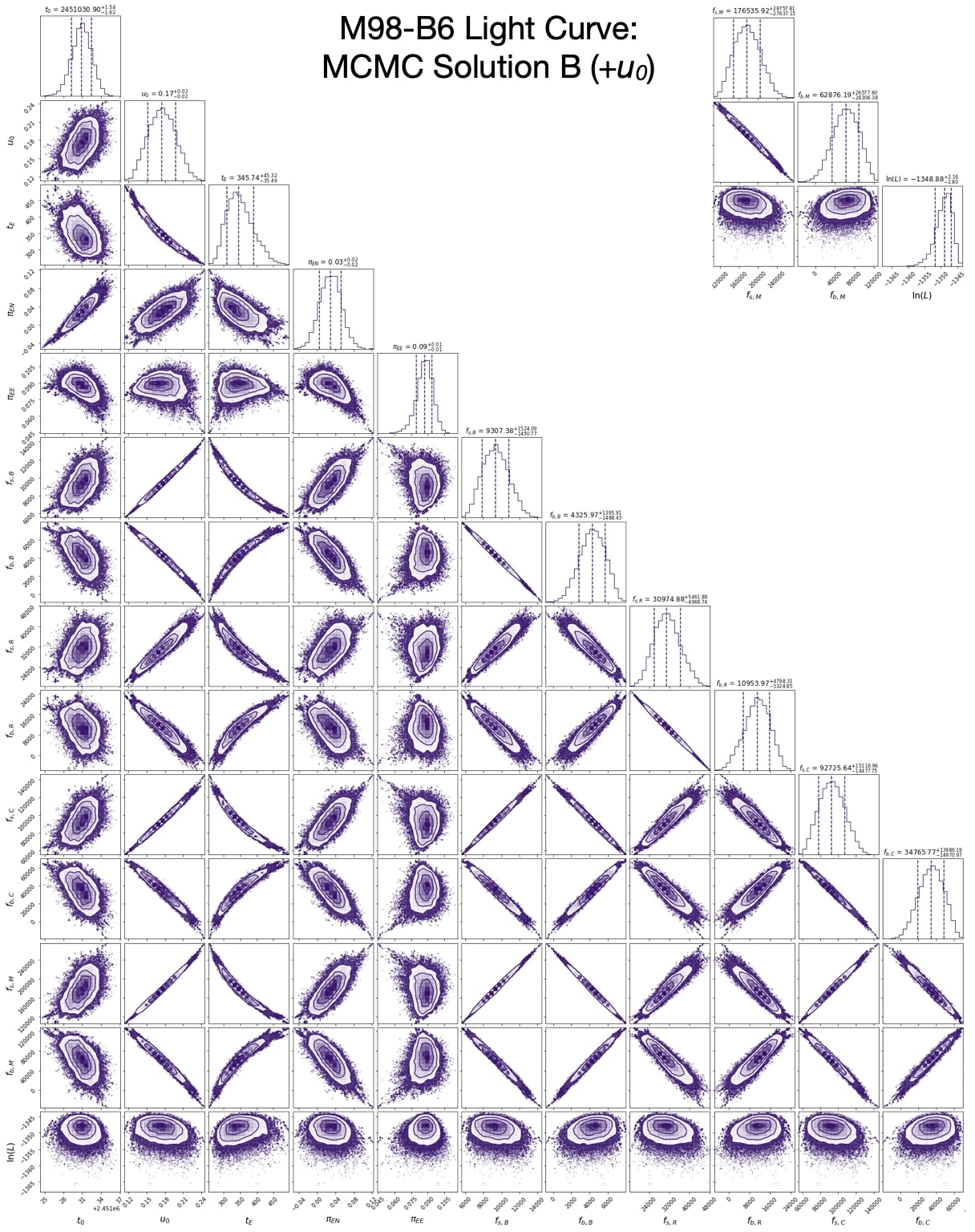}
	\caption{Posterior distributions from one of two MCMC solutions for the light curve of M98-B6, here with positive $u_{\textrm{0}}$. In addition to the 13 parameters defined for Table \ref{tab:fits}, there is an additional row showing the log-likelihood $\ln{L}$, defined here as $-0.5\chi^2$. Note that the rightmost three columns have been re-positioned to fit the figure space.}
\end{figure*}

\begin{figure*}[h!]
	\centering
	\label{cmds_app}
	\includegraphics[width=\textwidth]{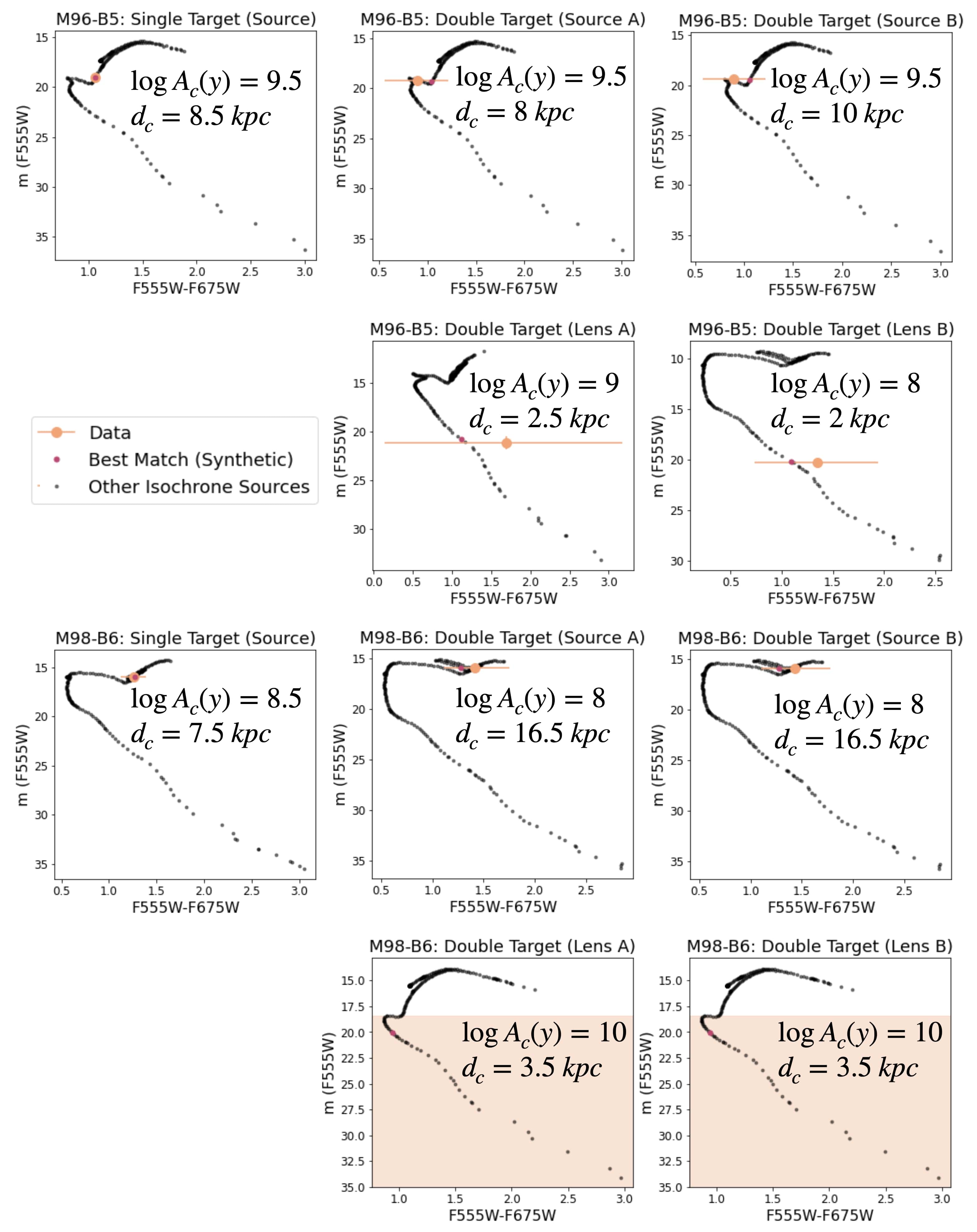}
	\caption{For each of the matched synthetic objects in Table \ref{tab:phot}, the isochrone from which that object was pulled is illustrated in a single panel here, with the matched source shown in magenta and all other sources in black. The corresponding observed data point is shown on each panel in orange with errorbars. For the two instances where we have only a limit on the object magnitude (both lenses for M98-B6), the range allowed by the data is shaded in orange. For each match, the log of the cluster age in years and the cluster distance in kpc is reported.}
\end{figure*}

\clearpage

\bibliography{references.bib}
\bibliographystyle{aasjournal}

\end{document}